\documentclass{article}
\usepackage[centertags]{amsmath}
\usepackage{hyperref}
\usepackage{color}
\usepackage{amsfonts}
\usepackage{amssymb}
\usepackage{amsthm}
\usepackage{cite}
\usepackage{fullpage}
\usepackage{graphicx}
\usepackage{pgf}
\usepackage{graphics}
\usepackage{tikz}
\usetikzlibrary{angles, calc, quotes}
\usetikzlibrary{shapes.geometric}
\usepackage{rotating}
\usepackage{asymptote}
\usetikzlibrary{math} 
\usepackage{bm}\usepackage{authblk}
\usetikzlibrary{decorations.pathreplacing,calligraphy,backgrounds}
\usetikzlibrary{decorations.shapes}
\usetikzlibrary{patterns}
\usepackage{mathtools}
\usetikzlibrary{intersections,decorations.markings}

\numberwithin{equation}{section}

\newcommand{\ii}{{\rm i}}
\newcommand{\dd}{{\rm d}}
\newcommand{\ee}{{\rm e}}

\title{One constant to rule them all}

\date{\vspace{-5ex}}
\author{}
 
\begin{document}

\maketitle

\vspace{6pt}
\begin{center}	
	{\textsl
	Aleksei Bykov\footnote{\scriptsize \tt aleksei.bykov@alumni.uniroma2.eu}},Ekaterina Sysoeva$^{\dagger}$\footnote{\scriptsize \tt sysoeva.caterina@gmail.com} \\
\vspace{1cm}
$\dagger$\textit{\small Universit\`a di Torino, Dipartimento di Fisica, Via P. Giuria 1, I-10125 Torino, Italy \\ I.N.F.N.- sezione di Torino, Via P. Giuria 1, I-10125 Torino, Italy }
\vspace{6pt}
\end{center}

\begin{center}
\textbf{Abstract}
\end{center}

\vspace{4pt}
{\small
    \par We study the coupling matrix of $\mathcal{N}=2$ $SU(N)$ gauge theories with $2N$ fundamental hypermultiplets in the special vacuum, where a residual $\mathbb{Z}_N$ symmetry restores nontrivial modular structure. Using symmetry and dimensional arguments, we construct its general form and identify $\lfloor N/2 \rfloor$ coupling constants in their most natural basis.

    \par We show that in the massless theory these couplings transform independently under $S$-duality and that the bare coupling is a modular function of any of them. One coupling constant, however, plays a distinguished role, emerging in the asymptotic regime and in instanton recursion relation. In the massive case, this structure is deformed but the distinguished coupling retains its privileged role.

}

\tableofcontents

    \section{Introduction}
    
       In their seminal works, Seiberg and Witten derived the exact low-energy effective action for $\mathcal{N}=2$ $SU(2)$ gauge theories, both with and without matter hypermultiplets, in terms of elliptic curves \cite{SW1, SW2}. Among these, theories whose matter content renders them conformal in the massless limit are of particular interest. They are related to two-dimensional conformal field theories via the AGT correspondence \cite{AGT} and exhibit nontrivial modular properties. In particular, in the conformal $SU(2)$ case, the theory is independent of any intrinsic scale and is characterised by a dimensionless parameter $\tau$, interpreted as the infrared gauge coupling. The bare coupling $\tau_{UV}$ is a modular function of $\tau$, while the coefficients of the Seiberg–Witten curve are modular forms under the natural action of $\Gamma(2)\subset SL(2,\mathbb{Z})$. 
        \par When nonzero masses are introduced for the multiplets, the modular properties are no longer manifest. Nevertheless, $S$-duality, which relates the weak- and strong-coupling regimes, imposes a modular anomaly equation on the prepotential. As a consequence, the prepotential can be expressed in terms of quasi-modular forms \cite{Pesandoetal, Pesandoetal2, Moralesetal}.
        \par The question of whether other theories, such as $\mathcal{N}=2^*$ theories with different gauge groups \cite{ADE, nonsimplaced} and the higher rank $SU(N)$ theories \cite{MinahanNemeschansky, Frauetal,Billoetal,Lerda}, possess similar modular properties or not attracts a fair amount of interest.   
        \par In this paper, we focus on $SU(N)$ theories with $2N$ fundamental hypermultiplets. Although the Seiberg-Witten curve for these theories can be constructed, for example, using equivariant localisation \cite{NekrasovOkunkov}, no clear modular structure emerges at generic points of the moduli space. An exception occurs at the so-called `special vacuum', where the vacuum expectation values of the Higgs field $a_u$ are arranged near the vertices of a regular $N$-polygon. In this configuration, a residual $\mathbb{Z}_N$ symmetry reduces the number of independent components of the coupling matrix \cite{ArgyresPelland}, and modular properties reappear.
        \par For $N=3$, the Seiberg–Witten curve in the special vacuum was constructed using these modular properties \cite{MinahanNemeschansky}, and the prepotential and coupling matrix were later obtained via localisation in \cite{Frauetal}, confirming the expected structure. In particular, the bare coupling is a modular function of the infrared coupling of the massless theory, while nonzero masses lead to quasi-modular corrections.
        \par For $N \geq 4$, the situation becomes more intricate. The effective coupling matrix is no longer proportional to the classical one, even in the special vacuum \cite{MinahanNemeschansky}. In the case $N=4$, it was shown in \cite{Billoetal} that the coupling matrix depends on two independent parameters, $\tau^{(1)}$ and $\tau^{(2)}$, and that the bare coupling can be expressed as a modular function of either of them.
        \par Finally, in \cite{Lerda} the structure of the coupling matrix in the special vacuum was considered for $N=2,\ldots,7$. The authors proposed a nontrivial algorithm for the construction of the coupling matrix based on an analysis of the perturbative contribution and the action of the $S$-duality group. They observed that the considered cases are parametrised by $\lfloor\frac{N}{2}\rfloor$ couplings, each constant transforming independently under the action of the $S$-duality group, and that the bare coupling constant is a modular form of any of them, and they conjectured that the same algorithm can be applied to general $N$. The authors studied the theories with massive multiplets and the consequent modular anomaly equation imposed on the prepotential as well.
        \par While this picture suggests that all these couplings play an equal role, we observed in our recent work \cite{BykovSysoeva} that in the asymptotic regime of the special vacuum, where the radius of the defining polygon becomes large, a single coupling constant emerges naturally as distinguished. This same constant appears in the recurrence relation governing the instanton partition function and thus plays a role not only in the special vacuum but across generic vacua, in both the massless and massive cases. In the $N=4$ theory, this distinguished coupling coincides, in an appropriate basis, with one of the two couplings identified in \cite{Billoetal}. Furthermore, an analysis of its associated monodromy group indicates that it corresponds to one of the $\lfloor N/2 \rfloor$ couplings introduced in \cite{Lerda} for massless flavours.  
        \par So why is it that all of $\lfloor\frac{N}{2}\rfloor$ constants are equally important, but one is more equally important than others?  In the present paper, we answer this question. We explicitly construct the coupling matrix in the special vacuum for arbitrary rank $N$, using dimensional and symmetry arguments, and thereby explain why a single coupling becomes distinguished in the asymptotic regime. We then find all $\lfloor\frac{N}{2}\rfloor$ coupling constants of the massless theory. These constants are numbered in a natural way, and the special constant appearing in the recurrence relation is the first one among them. We see that these constants indeed transform independently under $S$-duality as previously conjectured in \cite{Lerda} and that the bare coupling constant is a modular function of any of them. Furthermore, for the theory with the massive matter multiplets, we find that the coupling constants deformed by masses transform under $S$-duality \textit{almost} independently. Namely, they transform through themselves and the above-mentioned first of $\lfloor\frac{N}{2}\rfloor$ coupling constants, which thus retains a special role.
        
        \vspace{12 pt}
        \par The paper is organised as follows:
        \begin{itemize}
            \item In Section \ref{sec:specvac}, we recall what the special vacuum is and its specific symmetry and introduce symmetrical variables convenient for its study.
            \item In Section \ref{sec:generalform}, we derive the most general form of the coupling matrix based on the Weyl symmetry and dimensional reasons only.
            \item In Section \ref{sec:couplconst}, we find all the coupling constants of the theory
            \item and we study their modular properties in Section \ref{sec:modularproperties}.
            \item In Section \ref{sec:highord}, we show how the high order corrections to the SW curve can be  found.
            \item Finally, in Section \ref{sec:masscorrections}, we study how nonzero masses of the multiplets affect the coupling constants and modify their modular properties.
        \end{itemize}

    \section{Special vacuum} \label{sec:specvac}
    
        \subsection{Special vacuum and its symmetry}
            We will be working with the superconformal theory on $\mathbb{R}^4$ with $\mathcal{N}=2$ supersymmetry, the gauge group $SU(N)$ and $2N$ matter hypermultiplets in the fundamental representation.
            \par We consider the special vacuum of the theory and small deviations from it. We start with the massless case, when the special vacuum is simply defined by placement of the vacuum expectation values of the Higgs field $a_u$ at the vertices of a regular $N$-polygon.
            \par It will be convenient to use symmetric variables
            \begin{equation} \label{eqn:varw}
                w_k= \sum_{{i_1}<{i_2}<\ldots<{i_{k+1}}} a_{i_1} \ldots a_{i_{k+1}} , \qquad k=0, \ldots, N-1 . 
            \end{equation}
            In $SU(N)$ theory we always have $w_0=0$.
            \par Let us  introduce a polynomial
            \begin{equation} \label{eqn:P0}
                P_0(x)=\prod_{u=1}^N(x-a_u)=x^N+x^{N-2}w_1-x^{N-3} w_2 +\ldots - A^N,
            \end{equation}
            where $A$ is a complex constant such that
            \begin{equation}  \label{eqn:wtoAN}
                w_{N-1}=-(-A)^N .
            \end{equation}
            Note that $w_i=0$ for $i=1, \ldots N-2$ in the unperturbed special vacuum, and the polynomial has the form
            \begin{equation}
                P_0(x)=x^N-A^N ,
            \end{equation}
            so unperturbed $a_u^{(0)}$ are indeed placed at the vertices of a polygon
            \begin{equation} \label{eqn:au0}
                a_u^{(0)}=A \ee^{ \frac{2\pi \ii u}{N}}, \qquad u=1,  \ldots ,  N .
            \end{equation}
            One can note that the special vacuum respects a residual Weyl symmetry $\mathbb{Z}_N$: although there is a preferred way of labelling of $a_u$, the first element can be chosen freely. The shift in the labelling of $a_u$ is equivalent to choosing another root of (\ref{eqn:wtoAN}) to define $A$
            \begin{equation} \label{eqn:numerationshift}
                \ a_u\to a_{u+d} \quad \leftrightarrow \quad   A\to A \, \ee^{\frac{2\pi \ii d}{N}} .
            \end{equation}  
    
        \subsection{Perturbations of the special vacuum} \label{sec:pertutbation}
            \par Let us now allow small fluctuations $\delta a_u$ around the special vacuum that preserve the condition $\sum_u a_u =0$. We assume them to be of the same order, which we symbolically denote as $\delta a_{u} = \mathcal{O}( \delta)$ for any $u$.
            \par
            The symmetry (\ref{eqn:numerationshift}) indicates that the discrete Fourier transform with respect to the index $u$ may be useful. We define
                    \begin{equation}\label{eqn:a-to-v}
                v_l=\frac{1}{\sqrt{N}} \sum_{u=1}^N  \ee^{\frac{2\pi \ii l u}{N}}\delta a_{u},\ l=1,\ldots,N-1.
            \end{equation}
            Note that we can allow $l=0$, but then $v_0$ is always zero.
           Then
            \begin{equation} \label{eqn:v-to-a}
                \delta a_{u}=\frac{1}{\sqrt{N}}\sum_{l=1}^{N-1}  \ee^{-\frac{2\pi \ii l u}{N}}v_l.
            \end{equation}

            \par To see how the Fourier coefficients $v_l$ are related to the symmetric variables $w_k$, we introduce
            \begin{equation}
                w_l^{[u]}=\sum_{\substack{u_1<u_2<\ldots<u_{l+1}\\ u_1,u_2,\ldots,u_{l+1}\neq u}} a_{u_1}\cdots a_{u_{l+1}}.
            \end{equation}
            Then we can write
            \begin{equation}
                \delta w_{l}=\sum_{u=1}^N   w_{l-1}^{[u]} \delta a_{u}+\mathcal{O}(\delta^2).
            \end{equation}
            It is easy to see that
            \begin{equation}
                w_{l}^{[u]}=w_{l}-a_u w^{[u]}_{l-1}.
            \end{equation}
            This rule is, in fact, also correct for $l=0$ if we define $w_{-1}=1$. 
            Continuing this recursively, and taking into account  that for all $l=0,\ldots,N-2$ in the special vacuum $w_l$ vanish, we get
            \begin{equation}\label{eqn:dwi-is-dau}
                \delta w_{l}=(-1)^l \sum_{u=1}^N A^{l} \ee^{\frac{2\pi \ii l u}{N}}\delta a_{u}+\mathcal{O}(\delta^2),
            \end{equation}
            or
             \begin{equation}
                \delta w_l=\sqrt{N} v_l (-A)^l+\mathcal{O}(\delta^2). \label{eqn:deltaw=v}
            \end{equation}
            We note that $w_i=\delta w_i = \mathcal{O}(\delta)$ for $i=1, \ldots, N-2$. 
            %Then from (\ref{eqn:P0}) we find
            % \begin{equation}
            %     a_u=a_u^{(0)}+a_u^{(1)}+a_u^{(2)}+\mathcal{O}(\delta^3),
            % \end{equation}
            % where
            % \begin{equation} \label{eqn:deltaau1}
            %     a_u^{(1)}=\frac{1}{N}\sum_{t=1}^{N-2}(-1)^t \frac{w_t}{A^t} \ee^{-\frac{2\pi\ii u t}{N}} ,
            % \end{equation}
            % \begin{equation} \label{eqn:deltaau2}
            %     a_u^{(2)}=\frac{1}{N^2}\sum_{k=1}^{N-2}\sum_{t=1}^{N-2}\frac{w_k w_t}{A^{k+t+1}}\left(\frac{N-1}{2}-k \right) \ee^{-\frac{2 \pi \ii (k+t+1)}{N}}.
            % \end{equation}
            %One may notice that the power of $A$ appearing in the Fourier decomposition of $\delta a_u$ matches perfectly the harmonic, so, as expected, $\delta w_l$ is invariant with respect to the symmetry (\ref{eqn:numerationshift}).
            % \par Let us now introduce a Fourier transformation of the perturbation $\delta a_u$.
    
            % %%%
            % from (\ref{eqn:dwi-is-dau}) we have
            We shall also need the second-order correction for $\delta w_1$. It is given by
            \begin{equation}
                \delta w_1^{(2)}=\sum_{u<v}\delta a_u \delta a_v=\frac{1}{2}\sum_{u,v}\delta a_u \delta a_v-\frac{1}{2}\sum_u \delta a_u^2=-
                \frac{1}{2}\sum_{l=1}^{N-1}v_l v_{N-l},
                \end{equation}
                so
                \begin{equation}\label{eqn:w1exact}
                    w_1=-A \sqrt{N} v_1-
                \frac{1}{2}\sum_{l=1}^{N-1}v_l v_{N-l}.
                \end{equation}
                This is exact because $w_1$ is a quadratic polynomial in $a_u$, so no higher order corrections can appear.
                \par
            We also introduce the dual variables
            \begin{equation}
                v^D_l=\frac{\partial \mathcal{F}}{\partial v_l},
            \end{equation}
            where the partial derivative is taken with $v_j$, $j\neq l$ being fixed. We have
            \begin{equation}\label{eqn:wdi}
                v^D_l=\frac{1}{\sqrt{N}} \sum_{u=1}^N (\hat{\partial}_{a_u} \mathcal{F}) \ee^{-\frac{2\pi \ii l u}{N}}=\frac{1}{\sqrt{N}}\sum_{u=1}^N \hat{a}_u^{D} \ee^{-\frac{2\pi \ii l u}{N}},
            \end{equation}
            where $\hat{\partial}_{a_u}$ denotes a formal operation of differentiation with respect to $a_u$ with $a_v$ held fixed for $v=1,\ldots,N$, $v\neq u$.
            %\vspace{1 ex}
            %\noindent
            \par  Strictly speaking, the operator $\hat{\partial}_{a_u}$ cannot act on the function $\mathcal{F}$, defined on the hypersurface $a_1+\ldots+a_N=0$ only. As a consequence, the dual variables $\hat{a}_u^D$ cannot be rigorously defined by  $\hat{a}_u^D=\hat{\partial}_{a_u} \mathcal{F}$. One can formally define $\hat{a}_u^D$ in the Seiberg-Witten theory as an integral of the Seiberg-Witten differential over a certain contour, but this integral diverges \cite{NekrasovOkunkov}.  Fortunately, the divergent part does not depend on $u$, and thus does not contribute to $v_l^D$ for $l=1,\ldots,N-1$. 
            For convenience we can renormalise $\hat{a}_u^D$ as in \cite{NekrasovOkunkov}, by setting
            \begin{equation}
             v_0^D \triangleq \sum_{u=1}^N\hat{a}_u^D=0.   
            \end{equation} 
            With such a convention, we can invert the Fourier transform in (\ref{eqn:wdi}) and get
            \begin{equation}\label{eqn:hataud-v}
                \hat{a}_u^D=\frac{1}{\sqrt{N}}\sum_{l=1}^{N-1}v_l^D\ee^{\frac{2\pi \ii l u}{N}}.
            \end{equation}
            \par In this notation, the variables widely used in the literature can be written as
            \begin{equation}
                a_{u}^D=\frac{\partial{\mathcal{F}}}{\partial a_u}=\hat{a}_u^D-\hat{a}_N^D.
            \end{equation}
            Here the derivative is taken with respect to $a_u$ with $a_v$ held fixed for $v=1,\ldots,N-1$ and $v\neq u$, and keeping $\sum_{u=1}^{N}a_u=0$. Then
            \begin{equation}\label{eqn:wdi-N}
                v^D_l=\frac{1}{\sqrt{N}}\sum_{u=1}^{N-1} a_u^{D} \ee^{-\frac{2\pi \ii l u}{N}}.
            \end{equation}
            So, the dual variables $v_l^D$ can be understood as the Fourier coefficients of either $a_u^D$ or $\hat{a}_u^D$.
            \par 
            It is also convenient to write $v_l^D$ in terms of the differences $\hat{a}^D_{u-1}-\hat{a}^D_{u}$, which are clearly well-defined, with $\hat{a}^D_{0}=\hat{a}^D_{N}$. From (\ref{eqn:wdi}) it is easy to see that
            \begin{equation}\label{eqn:vdviaargcyc}
                v^D_l=\frac{1}{\sqrt{N}}\frac{1}{(1-\ee^{-\frac{2\pi \ii l }{N}})}\sum_{u=1}^{N} (\hat{a}_{u}^{D}-\hat{a}_{u-1}^D) \ee^{-\frac{2\pi \ii l u}{N}}
            \end{equation}
           \par  Note that $v_l$ and their duals transform under the residual Weyl symmetry (\ref{eqn:numerationshift}) as
            \begin{equation}\label{eqn:vtrans}
                v_l\to   \ee^{-\frac{2\pi \ii l d}{N}} v_l,\ v^D_l\to   \ee^{\frac{2\pi \ii l d}{N}} v_l^D.
            \end{equation}
            \par From (\ref{eqn:deltaw=v}), (\ref{eqn:numerationshift}) and (\ref{eqn:vtrans}), we see that $\delta w_l$ are invariants of the symmetry, as they should be.
            % \par  In other words, (\ref{eqn:a-to-v}) and (\ref{eqn:v-to-a}) can be understood as a decomposition of all possible deformations of the special vacuum into the eigenmodes of the residual Weyl symmetry. Equation (\ref{eqn:wdi}) has analogous sense in the dual space.

    \section{Coupling matrix} \label{sec:generalform}
        
        In this section we analyse the most general form of the coupling matrix based on how the prepotential can depend on the variables $v_l$ taking into account the dimensional and symmetric constraints.
        \par We assume that the special vacuum itself with $A\neq 0$ is a regular point of the prepotential. As a consequence, we expect that the prepotential in the vicinity of the special vacuum can be approximated by a sum of nonnegative powers of $v_l$. We are interested in terms up to the order $\delta^2$.
        \par First of all, the prepotential should have dimension two (assuming that $a_u$ are of dimension one). The variables $v_l$ are of dimension one, so up to $ \mathcal{O}(\delta^2)$ we can have terms of the form $A^2$, $v_l A$ and  $v_l v_{l'}$. 
        \par Secondly, the prepotential should be invariant with respect to the residual Weyl symmetry given by (\ref{eqn:numerationshift}), (\ref{eqn:vtrans}), so in the linear case we may have only $l=1$, while in the quadratic term $l+l'=N$. Besides, we see immediately that for $N > 2$ the term $A^2$ is not possible. We will now consider $N>2$ and see later that the term $A^2$, possible for $N=2$, does not introduce anything new. Then the most general expansion of the prepotential is
        \begin{equation}\label{eqn:FviaV}
            \mathcal{F}=\mathcal{F}_0+{2\pi\ii}\left( -{\tau}_{\mathrm{IR}}\sqrt{N} v_1 A-\frac{1}{2}\sum_{l=1}^{N-1}v_{l}v_{N-l} \tau^{(l)}\right)+\mathcal{O}(\delta^3)
        \end{equation}
        with $\tau^{(l)}=\tau^{(N-l)}$.
        \par In (\ref{eqn:FviaV}) we set 
        \begin{equation}
        {\tau}_{\mathrm{IR}}=-\frac{1}{\sqrt{N}A}\frac{1}{2\pi\ii}\frac{\partial\mathcal{F}}{\partial v_1}=\frac{1}{2\pi\ii}\frac{\partial\mathcal{F}}{\partial w_1}    
        \end{equation}
        which is consistent with what we found for the unique coupling constant appearing in the asymptotic behaviour of the partition function in \cite{BykovSysoeva}.
        \par Alternatively, we can interpret this coefficient as a background value of the dual variable $v_1^D$,
        \begin{equation}
            v_{l}^{D(0)}=\left(\frac{\partial\mathcal{F}}{\partial v_l}\right)_{v=0}=-2\pi \ii A {\tau}_{\mathrm{IR}}\delta_{l,1}.
        \end{equation}
        % Given that for $l=2,\ldots,N-1$ the dual variable $v_l^D$ has no background value at the special vacuum, imposing by hands a condition $\sum_{u}a_u^{D(0)}=0$ as in \cite{NekrasovOkunkov} to get rid of irrelevant divergent constant, and reversing the Fourier transform in (\ref{eqn:wdi}), we get
        Then, by (\ref{eqn:hataud-v})
        \begin{equation}\label{eqn:auD0}
            \hat{a}_u^{D(0)}=-2\pi\sqrt{N} \ii A {\tau}_{\mathrm{IR}} \ee^{\frac{2\pi\ii u}{N}}=-\frac{2\pi \ii}{\sqrt{N}}  {\tau}_{\mathrm{IR}} a_u.
        \end{equation}    
        \par 
        Let us see how the coefficients $\tau^{(l)}$ are related to the coupling matrix
        \begin{equation}
            \tau_{uv}^{\mathrm{IR}}=\frac{1}{2 \pi \ii}\frac{\partial^2 \mathcal{F}}{\partial a_{u} \partial a_{v}}.
        \end{equation}
        By (\ref{eqn:FviaV}) and (\ref{eqn:a-to-v}) we get
        \begin{equation}
            \tau^{(l)}=-\frac{1}{2\pi\ii}\frac{\partial^2 \mathcal{F}}{\partial v_{l} \partial v_{N-l}}=-\frac{1}{2\pi\ii N}\sum_{u,v}\frac{\partial^2 \mathcal{F}}{\partial a_{u} \partial a_{v}}\ee^{\frac{2\pi\ii (u-v)l}{N}}=-\frac{1}{ N}\sum_{u,v}\tau_{uv}^{\mathrm{IR}}\ee^{\frac{2\pi\ii (u-v)l}{N}}.
        \end{equation}  
        Furthermore, since
        \begin{equation}
            v_{N-1}=-(-1)^N\frac{\delta w_{N-1}}{\sqrt{N}A^{N-1}}+O(\delta^2)=\sqrt{N}\delta A+\mathcal{O}(\delta^2),
        \end{equation}
        and the prepotential $\mathcal{F}$ cannot depend individually on $A$ or $\delta A$, but only on their sum, we conclude that $\tau_{\mathrm{IR}}=\tau^{(1)}$. For the same reason, the term with $A^2$, possible for the $N=2$ theory, cannot bring any new coupling constant in the picture. 
        \par 
    %    The first term can be combined with the rest. For that we  get rid of this redundancy by setting $v_{N-1}=0$ and use $w_i$ instead of $v_l$. We should be careful with the first term, because (\ref{eqn:deltaw=v}) holds only up to $\mathcal{O}(\varepsilon^2)$. The precise equation is
        For further convenience, let us write the prepotential in terms of the variables $w_k$. We use (\ref{eqn:deltaw=v}) for $k>1$ and (\ref{eqn:w1exact}) for $w_1$, because the latter enters linearly.
        We have:    
        % \begin{equation}
        %     w_1+A\sqrt{N}v_1=\sum_{u<v}\delta a_u \delta a_v=\frac{1}{2}\left(\sum_{u}\delta a_u\right)^2-\frac{1}{2}\sum_{u}\delta a_u^2=-\frac{1}{2}\sum_{l=1}^{N-1}v_l v_{N-l}.
        % \end{equation}
        % Using this together with $w_l=\delta w_l$ for $l=2,\ldots,N-2$, we get
    %    \begin{eqnarray}\label{eqn:genF}
    %        \mathcal{F}={2\pi\ii}\left(\tau^{\mathrm{IR}}_1 w_1-\frac{1}{2}\sum_{l=2}^{N-2}v_ {l}v_{N-l}   \left(\tau^{(l)}-\tau^{\mathrm{IR}}_1\right)\right)+\mathcal{O}(\varepsilon^3)= \pi\ii\sum_{l=1}^{N-1}\frac{w_{l} w_{N-l}}{w_{N-1}} \hat{\tau}_l^{\mathrm{IR}} +\mathcal{O}(\varepsilon^3),
    %      \end{eqnarray}
        \begin{eqnarray} \label{eqn:genF}
            \mathcal{F}=2\pi \ii \left(\tau_{\rm IR} w_1 -\frac{1}{2N}\sum_{l=2}^{N-2}\frac{w_{l} w_{N-l}}{w_{N-1}}(\tau_{\rm IR}-\tau^{(l)})\right)=\frac{2 \pi \ii}{2N}\sum_{l=1}^{N-1}\frac{w_{l} w_{N-l}}{w_{N-1}}\hat{\tau}^{(l)},
        \end{eqnarray}
        where 
        \begin{equation}
            \hat{\tau}^{(l)}=\tau^{(l)}-\tau_{\rm IR}(1-2N \delta_{l,1}).
        \end{equation}
    %    where
    %     \begin{equation}
    %      \hat{\tau}_1^{\mathrm{IR}}=\hat{\tau}_{N-1}^{\mathrm{IR}}={\tau}_1^{\mathrm{IR}}; \qquad\hat{\tau}_l^{\mathrm{IR}}={\tau}_l^{\mathrm{IR}}-{\tau}_1^{\mathrm{IR}}\ \mathrm{for}\ l=2,\ldots,N-2.
    %    \end{equation}
    %    In this form all the coupling constants $\tau^{(l)}$ enter on equal footing, in line with the ideology of \cite{Lerda}. However, it is important to underline that the contribution $l=1,N-1$ is special in two ways.
        We now see that $\tau_{\mathrm{IR}}=\tau^{(1)}$ is special for two reasons. Firstly, only this term brings a nontrivial background value for the dual variables (\ref{eqn:auD0}).  Secondly, from (\ref{eqn:genF}) it is clear that this is always the only contribution that survives in the limit of large $w_{N-1}$ studied in \cite{BykovSysoeva}.
        \par
        From (\ref{eqn:genF}) we see that the classical part of the prepotential $\mathcal{F}_{\mathrm{class}}=w_1 \ln(q_0)$ is given by setting 
        \begin{equation}\label{eqn:tau-class}
            \tau^{(l)}_{\mathrm{class}}=\frac{1}{2\pi \ii}\ln(q_0),\ l=1,\ldots,N-1.
        \end{equation}
        Here we denote the bare coupling constant by $q_0$ to be consistent with \cite{BykovSysoeva}. In our notation $q=(-1)^N q_0$.
        \par 
        Let us now compute the full coupling matrix. It is convenient to start from (\ref{eqn:FviaV}), where the linear and quadratic parts are separated. We have
        \begin{eqnarray} \label{eqn:tau-decomose}
           \tau_{uv}^{\mathrm{IR}}&=&\frac{1}{2\pi\ii}\frac{\partial^2 \mathcal{F}}{\partial a_u \partial a_v}=\frac{1}{2\pi\ii}(\hat{\partial}_{a_u}-\hat{\partial}_{a_N})(\hat{\partial}_{a_v}-\hat{\partial}_{a_N})  \mathcal{F} =\frac{1}{ N}\sum_{l=1}^{N-1}\tau^{(l)}\left(\ee^{\frac{2\pi\ii u l}{N}}-1\right)\left(\ee^{-\frac{2\pi\ii v l}{N}}-1\right) \nonumber \\ &=& 
            \frac{1}{ N}\sum_{l=1}^{N-1}\tau^{(l)}    \left(\cos\left(\frac{2\pi (u-v)l}{N}\right)-\cos\left(\frac{2\pi u l}{N}\right)-\cos\left(\frac{2\pi v l}{N}\right)+1\right).
        \end{eqnarray}
        \par This is the most general form of the coupling matrix, not violating the residual Weyl symmetry. Note that it depends on $\lfloor\frac{N}{2}\rfloor$ independent coefficients as expected. Only one of them, $\tau^{(1)}$, can be seen in the large $w_{N-1}$ asymptotic, while the others contribute to the next orders only. In the cases $N=2,3,4,5,6,7$, this decomposition coincides with the one presented in \cite{Lerda}. We note that in the cited work the basis of coupling matrices was built using an explicit, yet nontrivial algorithm starting with the one-loop coupling matrix. Instead, we have found a simple explicit expression for any rank $N$ based on symmetry and dimensional principles only.
        \par 
        In agreement with observations made in \cite{Lerda}, the sum of all basis coupling matrices in (\ref{eqn:tau-decomose}) gives the classical coupling matrix,
        \begin{eqnarray}
            &&\frac{1}{N}\sum_{l=1}^{N-1}\left(\cos\left(\frac{2\pi (u-v)l}{N}\right)-\cos\left(\frac{2\pi u l}{N}\right)-\cos\left(\frac{2\pi v l}{N}\right)+1\right)\\
            &&=  \delta_{uv}+1\ (\mathrm{for}\ u,v=1,\ldots,N-1).
        \end{eqnarray}
        This is another way to state that the classical prepotential is realised by setting all the coupling constants to the same value as in (\ref{eqn:tau-class}).

    \section{Coupling constants} \label{sec:couplconst}
    
        Let us now find explicitly all the $\lfloor\frac{N}{2} \rfloor$ coupling constants appearing in the coupling matrix (\ref{eqn:tau-decomose}).
    
        \subsection{Perturbative part of the prepotential}
            \paragraph{Classical contribution.} As we have already discussed, the classical contribution is given by
            \begin{equation} \label{eqn:tauclass}
                2 \pi \ii \tau^{(l)}_{\mathrm{class}}=\ln(q_0), \qquad l=1,\ldots,N-1.
            \end{equation}
        
            \paragraph{One-loop contribution.} The one-loop contribution to the coupling matrix is well known and can be found, for example, in \cite{MinahanNemeschansky} or \cite{Lerda}. However, to get the right imaginary part, we have to choose the branch of the logarithm properly. Namely, we have to choose it in such a way that the residual Weyl symmetry is preserved. For this we find it easier to start with the explicitly Weyl-symmetric one-loop prepotential presented in \cite{Nekrasov} (with corrected misprints)
            \begin{equation}
                \mathcal{F}_{\rm 1loop}=\frac{1}{2}\sum_{\substack{u,v=1 \\  u \neq v}}^N (a_u-a_v)^2 \ln \frac{a_u-a_v}{\Lambda} - N \sum_{u=1}^N a_u^2 \ln \frac{a_u}{\Lambda}.
            \end{equation}
            Then, for the second derivatives of the prepotential we get 
            \begin{equation}
              \hat{\partial}_u\hat{\partial}_v \mathcal{F}_{\rm 1loop}= \delta_{u,v} \left(\sum_{\substack{k=1 \\ k \neq u}}^N \ln \left(-\frac{(a_{u}-a_k)^2}{a_u^2} \right)-\ln a_u^2  \right)+(1-\delta_{u,v})(-\ln(-(a_{u}-a_v)^2)) + \rm{const} ,
            \end{equation}
            where the last term denoted as $\rm{const}$ does not depend on $a_u$ and will not play any role.
            \par With (\ref{eqn:v-to-a}) for the deviation of the prepotential we get
            \begin{eqnarray}
                &&\delta\mathcal{F}_{\rm 1loop}=\frac{1}{2}\sum_{u,v=1}^N \delta a_u \delta a_v \hat{\partial}_u\hat{\partial}_v \mathcal{F}_{\rm 1loop}= \frac{1}{2N}\sum_{m,l=1}^{N-1}v_l v_m \sum_{u,v=1}^N \ee^{-\frac{2 \pi \ii u l}{N}}\ee^{-\frac{2 \pi \ii v m}{N}} \cdot \\
                && \left( \delta_{u,v} \left(\sum_{\substack{k=1 \\ k \neq u}}^N \ln \left(-(1-\ee^{\frac{2\pi\ii (k-u)}{N}})^2 \right)-\ln \ee^{\frac{4 \pi \ii u}{N}}  \right)+(1-\delta_{u,v})(-\ln(-( \ee^{\frac{2 \pi \ii u}{N}}- \ee^{\frac{2 \pi \ii v}{N}})^2)) + {\rm const} \right). \nonumber
            \end{eqnarray}
            Since the Fourier transform of a constant is zero, the last term disappears. Besides,
            \begin{equation}
                \sum_{u,v=1}^N \ee^{-\frac{2 \pi \ii u l}{N}}\ee^{-\frac{2 \pi \ii v m}{N}} \left(\delta_{u,v} \frac{4 \pi \ii u}{N}+(1-\delta_{u,v}) \frac{2 \pi \ii (u+v)}{N}\right)= \sum_{u,v=1}^N \ee^{-\frac{2 \pi \ii u l}{N}}\ee^{-\frac{2 \pi \ii v m}{N}}  \frac{2 \pi \ii (u+v)}{N}=0.
            \end{equation}
            So the variation of the prepotential turns into
                
            \begin{eqnarray}
                &&\delta\mathcal{F}_{\rm 1loop}= \frac{1}{2N}\sum_{m,l=1}^{N-1}v_l v_m \cdot \\
                && \Big(\sum_{u}^N \ee^{-\frac{2 \pi \ii u (l+m)}{N}}\left(\sum_{k=1}^{N-1} \ln \left(4\sin^2\frac{\pi k}{N} \right)+ \pi \ii (N-1)  \right)-  \sum_{\substack{u,v=1 \\ u\neq v}}^N \ee^{-\frac{2 \pi \ii (u l+mv)}{N}}\ln\left(4\sin^2\frac{\pi(u-v)}{N}\right) \Big) \nonumber
            \end{eqnarray}
            Now, it is easy to see that the sums over $u$, $v$ vanish unless $l+m=N$.
            Moreover,
            \begin{equation}
                \sum_{k=1}^{N-1} \ln \left(4\sin^2 \frac{\pi k}{N} \right) = 2 \ln N ,
            \end{equation}
            and we arrive at
            \begin{eqnarray}
                &&\delta\mathcal{F}_{\rm 1loop}=  \frac{1}{2} \sum_{m=1}^{N-1}v_m v_{N-m} \Big( 2 \ln 2 N+\pi \ii (N-1)  -  \sum_{k=1}^{N-1} \ee^{-\frac{2 \pi \ii k m}{N}}\ln(\sin^2\frac{\pi k}{N}) \Big)  \\
                 &&= \frac{1}{2}\sum_{m=1}^{N-1}v_m v_{N-m} \Bigg( 2 \ln 2 N+ \pi \ii (N-1)  -  \sum_{k=1}^{N-1} \cos\left(\frac{2 \pi  k m}{N}\right)\ln\left(\sin^2\frac{\pi k}{N}\right) \Bigg).
            \end{eqnarray}
            Comparing this with (\ref{eqn:FviaV}) 
            and taking into account the already found classical contribution (\ref{eqn:tauclass}), we get the perturbative contribution to the coupling constants as
            \begin{equation}
                2\pi\ii \tau^{(l)}_{\mathrm{pert}}=\ln(q_0)- 2 \ln 2 N- \pi \ii (N-1)  +  \sum_{k=1}^{N-1} \cos\left(\frac{2 \pi  k l}{N}\right)\ln\left(\sin^2\frac{\pi k}{N}\right). 
            \end{equation}
            By rewriting it in terms of $q=(-1)^N q_0$ and choosing the branch of the logarithm as 
            \begin{equation}
                \ln(q)=\ln(q_0)-\pi \ii N ,
            \end{equation}
            we obtain
            \begin{equation}
                2\pi\ii \tau^{(l)}_{\mathrm{pert}}=\ln(q)- 2 \ln 2 N +  \sum_{k=1}^{N-1} \cos\left(\frac{2 \pi  k l}{N}\right)\ln\left(\sin^2\frac{\pi k}{N}\right)+\pi \ii. 
            \end{equation}
            %%We will denote
            %\begin{equation} \label{eqn:logD}
            %    \ln(D_m)=2 \ln 2 N -  \sum_{k=1}^{N-1} \cos(\frac{2 \pi   k m}{N})\ln(\sin^2\frac{\pi k }{N})+\pi \ii.
            %\end{equation}
            We note that by the Gauss digamma theorem for general $N$
            \begin{equation}\label{eqn:logD-l}
                2\pi\ii \tau^{(l)}_{\mathrm{pert}}=\ln(q)+2\psi\left(\frac{l}{N}\right)+2\gamma_E+\pi\cot\left(\frac{\pi l}{N}\right) 
            \end{equation}
            depends on the ratio $l/N$ only.
    
        \subsection{Seiberg-Witten curve}
            To describe the instanton contribution, we will look for the effective number of instantons defined by
             \begin{equation}\label{keff-def}
               \frac{\partial \mathcal{F}_{\rm inst}(q)}{\partial \ln q}=\epsilon_1 \epsilon_2 k_{\rm eff}(q) .
            \end{equation}
            Given $k_{\rm eff}$ and the boundary condition $\mathcal{F}^{\rm inst}|_{q=0}=0$, we can reconstruct the prepotential.
            \par Since we are working with a nonequivariant problem, it is convenient to use the Seiberg-Witten curves \cite{SeibergWitten}. The equation for the curve can be derived from the Nekrasov partition function \cite{NekrasovOkunkov}. For an auxiliary function $\omega(y)$, following notation of \cite{BykovSysoeva}, in the massless case we can write an equation     
            \begin{equation} \label{eqn:structure}
                \omega\left(y\right) +\frac{1}{\omega(y)}  = \kappa\frac{P_N(y) }{y^N} ,
            \end{equation}
            where    
            \begin{equation}
                \kappa=\sqrt{q}+\frac{1}{\sqrt{q}},
            \end{equation}
            \begin{equation}
                P_N(y)= y^N-p_0 y^{N-1}+p_1 y^{N-2}+\ldots+(-1)^N p_{N-1}.
            \end{equation}
           % \begin{equation} \label{symQoly}
           %     Q(x)=\prod_{i=1}^N(x-m_i)=x^N-T_0 x^{N-1}+T_1 x^{N-2}+\ldots+(-1)^N T_{N-1}.
           % \end{equation}
            \par The periods of the function $\omega(y)$ are known to be
            \begin{equation} \label{eqn:Au-and-au}
                \oint_{\mathcal{A}_u} y \frac{\dd \omega(y)}{\omega(y)}= -2\pi \ii \frac{a_u}{A},
            \end{equation}
            where the contour $\mathcal{A}_u$ encircles the cut $\mathcal{C}_u$ of the function $\omega(y)$ such that the singular point $a_u$ belongs to the cut.
            \par The effective number of instantons appears in the asymptotic behaviour of $\omega(y)$ at $y \to \infty$ (see \cite{BykovSysoeva} for details).
            \begin{eqnarray} \label{eqn:omegaexp}
                \omega(y)=\sqrt{q}\frac{1}{1+\frac{w_1}{A^2y^2}}\left( 1-\frac{\epsilon_1\epsilon_1}{A^2y^2} k_{\rm eff}\right)+\mathcal{O}\left(\frac{1}{y^3}\right)= 
                \sqrt{q}\left(1-\frac{w_1+k_{\rm eff} \, \epsilon_1 \epsilon_2}{A^2y^2} \right)+\mathcal{O}\left(\frac{1}{y^3}\right). 
            \end{eqnarray}
            \par Comparing (\ref{eqn:structure}) and (\ref{eqn:omegaexp}) we conclude, that the first two coefficients of the polynomial $P_N$ are given by
            \begin{equation}\label{eqn:p0}
                p_0=0,
            \end{equation}
            \begin{equation}\label{eqn:p1-keff}
                p_1=\frac{(w_1+k_{\rm eff} \epsilon_1 \epsilon_2)(1-q)}{A^2(1+q)}.
            \end{equation}   
            \par Equations (\ref{eqn:structure}), (\ref{eqn:Au-and-au}) and (\ref{eqn:p1-keff}) are enough to find $k_{\rm eff}$, and hence the prepotential and the coupling matrix.
            \par In order to perform the integration in (\ref{eqn:Au-and-au}) we proceed in the same way as in \cite{BykovSysoeva}. First, we change the variables: we now consider $z={\omega}(y)$ as a new variable. Then $y$ becomes a function of $z$ defined by the equation
            \begin{equation}\label{eqn:magic-period}
                 {\omega}(y(z))=z.
            \end{equation}
            The multivalued function $y(z)$ in the region of interest has $N$ branches, which we denote by $y_u$ ($u=1,\ldots,N$), and all the contours $\mathcal{A}_u$ can be defined as the images of the same contour $\mathcal{Z}$ under the functions $y_u$, so instead of (\ref{eqn:Au-and-au}) we get
            \begin{equation} \label{eqn:periodsdz}
                \oint_{\mathcal{Z}}y_u(z) \frac{\dd z}{z}=-2\pi \ii \frac{a_u}{A},
            \end{equation}
            where the contour $\mathcal{Z}$ is to be determined. We note that, due to (\ref{eqn:au0}) and (\ref{eqn:v-to-a}), in terms of the variables $v_l$ the equation above takes the form
             \begin{equation} \label{eqn:periodsdz-v}
                \oint_{\mathcal{Z}}y_u(z) \frac{\dd z}{z}=-2\pi \ii \sum_{l=1}^{N-1}e^{-\frac{2\pi\ii l u}{N}}\left(\delta_{l,N-1}+\frac{v_l}{A\sqrt{N}}\right).
            \end{equation}
            \par Note that the function $\omega(y)$ does not respect the residual Weyl symmetry (\ref{eqn:numerationshift}). Instead, it is invariant under simultaneous transformations of $A$ and $y$
            \begin{equation} \label{eqn:Atrans-omega}
                \begin{cases}
                    A\to A \ee^{\frac{2\pi \ii d}{N}} ,    \\
                    y \to y \ee^{-\frac{2\pi\ii d}{N}} .
                \end{cases}
            \end{equation}
            \par Since the left-hand side of (\ref{eqn:structure}) is invariant under (\ref{eqn:Atrans-omega}), the right-hand side should be too. While the terms $y^N$ and constant $p_{N-1}$ of the polynomial $P_N$ are automatically invariant under (\ref{eqn:Atrans-omega}), the rest of the terms should have a corresponding power of $A$ to compensate their transformations, so the coefficients $p_k$ should be of the form
            \begin{equation}
                p_k y^{N-1-k} \sim \frac{y^{N-1-k}}{A^{1+k+t N}} ,
            \end{equation}
            where $t$ is an integer. But the parameter $A$ is dimensional, while $p_k$ is dimensionless, hence the dimensions of $A$ should be compensated by only dimensional parameters left, namely $w_m$. We conclude that $p_k$ for $k=1,\ldots,N-2$ are at least of order $\mathcal{O}(\delta)$, and therefore, in the leading order, equation (\ref{eqn:structure}) leads to the equation for the curve $y(z)$
            \begin{equation} \label{eqn:leadingorder}
                 \left(\frac{\left(z+\frac{1}{z}\right)}{\kappa}-1 \right)(y^{(0)}(z))^N +U^N=0 ,
            \end{equation}
            where $U=p_{N-1}$. 
              There are $N$ solutions of (\ref{eqn:leadingorder}), which can be written as
            \begin{equation}\label{eqn:y-cut}
                y_u^{(0)}(z)=U \rho(z)^{-1/N}\ee^{\frac{2 \pi u \ii}{ N}},\quad \rho(z)=\left(1-\frac{\left(z+\frac{1}{z}\right)}{\kappa}\right)
            \end{equation}
            with $u=1,\ldots,N$. 
            \par Equation (\ref{eqn:leadingorder}) is the same equation as we had in the leading order in the limit of $A \to \infty$ considered in \cite{BykovSysoeva}. We see now that the contour $\mathcal{Z}$ can be taken the same as in \cite{BykovSysoeva}.
            \begin{equation}
                \mathcal{Z}=\big\{z\in\mathbb{C}\big||z|=r\big\}, \quad r_0<r<1 ,
            \end{equation}
            where $r_0$ is such that solution (\ref{eqn:y-cut}) for sure have an analytical continuation in the region of interest. We can define it as
            \begin{equation*}
                \left|z+\frac{1}{z}\right|<|\kappa|,\quad \forall z: r_0<|z|<1 .
            \end{equation*}
            \par Let us define a function
            \begin{equation}
               2\pi\ii \, \mathcal{I}_{\alpha}(\kappa)=\oint_{|z|=r}\left(1-\frac{1}{\kappa} \left(z+\frac{1}{z}\right)\right)^{\alpha}\frac{dz}{z}.
            \end{equation}
            It can be expressed in terms of the hypergeometric function
            \begin{equation}
                \mathcal{I_{\alpha}}(\kappa)={}_2 F_1\left(-\frac{\alpha}{2},\frac{1-\alpha}{2};1;\frac{4}{\kappa^2}\right).
            \end{equation}  
            \par From the leading order of (\ref{eqn:periodsdz}) we find $U$ to be
            \begin{equation}\label{eqn:U-solvec}
                U=-\frac{1}{ \mathcal{I}_{-\frac{1}{N}}(\kappa)}=-\frac{1}{{}_2 F_1(\frac{1}{2N},\frac{N+1}{2N},1;\frac{4}{\kappa^2})}. 
            \end{equation} 
            \par Let us now consider the first subleading order of (\ref{eqn:structure}) with respect to small $\delta$ and determine the first correction to the curve.
            \begin{equation}\label{eqn:pk-zero}
               -\rho(z) N (y^{(0)})^{N-1}y^{(1)} +\sum_{k=0}^{N-2}(-1)^k p_k (y^{(0)})^{N-1-k}=0 ,
            \end{equation}
            so
            \begin{equation} \label{eqn:deltay1}
                y^{(1)}_u=\frac{\sum_{k=0}^{N-2}(-1)^k p_k (y^{(0)}_u)^{-k}}{\rho(z) N }.
            \end{equation}
            We substitute (\ref{eqn:deltay1}) in the left-hand side of (\ref{eqn:periodsdz-v}).
            Comparing the Fourier harmonics and using (\ref{eqn:deltaw=v}) we find that
            \begin{eqnarray}
                p_k^{(0)}=U^k \frac{w_k}{A^{k+1}} \frac{(-1)}{\mathcal{I}_{\frac{k}{N}-1}(\kappa)} +\mathcal{O}(\delta^2) .\label{eqn:pkleadingmassless}
            \end{eqnarray}
            In particular,
            \begin{equation}
                p_1^{(0)}= \frac{w_1}{A^{2}} \frac{1}{\mathcal{I}_{-\frac{1}{N}}(\kappa)\mathcal{I}_{\frac{1}{N}-1}(\kappa)} +\mathcal{O}(\delta^2).
             \end{equation}
            From (\ref{eqn:p1-keff}) we see that
            \begin{equation}
                \epsilon_1 \epsilon_2 k_{\rm eff}= p_1 A^2\frac{1+q}{1-q}-w_1= \left(-1 +\frac{1+q}{1-q} \frac{1}{\mathcal{I}_{\frac{1}{N}-1}\mathcal{I}_{-\frac{1}{N}}} \right)w_1 +\mathcal{O}(\delta^2).
            \end{equation}
            As we see, it coincides with the effective number of instantons found in \cite{BykovSysoeva} in $A \to \infty$ limit. It implies that the only full (\textit{i.e.}, including the perturbative contribution) coupling constant visible in the $A \to \infty$ limit satisfies the equation
            \begin{equation}\label{eqn:tauIRder}
                2\pi \ii q  \frac{ \partial \tau^{(1)}}{\partial q} =\frac{1+q}{1-q}\frac{1}{ \mathcal{I}_{-\frac{1}{N}}(\kappa)\mathcal{I}_{\frac{1}{N}-1}(\kappa)}=\frac{1+q}{1-q} \,{}_2F_1\left(\frac{1}{2N},\frac{N+1}{2N},1;\frac{4}{\kappa^2}\right)^{-2}\left(1-\frac{4}{\kappa^2}\right)^{\frac{N+2}{2N}}.
            \end{equation}
            Together with the boundary condition
            \begin{equation}\label{eqn:tauD}
                2\pi \ii \tau^{(1)}=2\pi\ii \tau^{(1)}_{\mathrm{pert}}+\mathcal{O}(q),\quad q\to 0,
            \end{equation}
            this yields the solution
            \begin{equation}\label{eqn:tau-1}
                2\pi \ii \tau^{(1)}=-\frac{\Gamma\left(\frac{1}{N}\right)^2}{\Gamma\left(\frac{2}{N}\right)}\frac{{{}_2F_1\left(\frac{1}{N},\frac{1}{N},\frac{2}{N},1-q\right)}}{{}_2F_1\left(\frac{1}{N},\frac{1}{N},1,q\right)}+\pi\left(\cot\left(\frac{\pi}{N}\right) +\ii\right).
            \end{equation}     
            \par We now proceed to consider the next order of (\ref{eqn:periodsdz}) to find all the other $\tau^{(l)}$.
            \par The second subleading order of (\ref{eqn:structure}) allows us to find the next correction to the curve.
            \begin{eqnarray}
                 &&-\rho(z) (N (y^{(0)})^{N-1}y^{(2)} +\frac{N(N-1)}{2} (y^{(0)})^{N-2}(y^{(1)})^2) \nonumber \\
                 &&+\sum_{k=0}^{N-2}(-1)^k p_k^{(0)} (y^{(0)})^{N-2-k}(N-1-k)y^{(1)}+\sum_{k=0}^{N-2}(-1)^k p_k^{(1)} (y^{(0)})^{N-1-k}=0 ,
            \end{eqnarray}
            so
            \begin{eqnarray}
                &&y^{(2)}=\frac{\sum_{k=0}^{N-2}(-1)^k p_k^{(0)} (y^{(0)})^{-1-k}(N-1-k)y^{(1)}+\sum_{k=0}^{N-2}(-1)^k p_k^{(1)} (y^{(0)})^{-k}}{\rho(z) N } \nonumber\\
                &&-\frac{(N-1)}{2}\frac{(y^{(1)})^2}{y^{(0)}}  . \label{eqn:deltay2}
            \end{eqnarray}
            Again, we substitute (\ref{eqn:deltay2}) in the left-hand side of (\ref{eqn:periodsdz}), while the right-hand side was already cancelled by the first order. Comparing the Fourier harmonics in the left- and right-hand sides of (\ref{eqn:periodsdz}) and using (\ref{eqn:pkleadingmassless}) we find
            \begin{eqnarray}
               &&p_1^{(1)}=\frac{(-1)^N}{N A^{N+2}}\sum_{k=2}^{N-2}w_k w_{N-k}\left(\frac{N-1}{2}-k \right)\left(\frac{1}{\mathcal{I}_{\frac{k}{N}-1}\mathcal{I}_{-\frac{k}{N}}}-\frac{1}{\mathcal{I}_{\frac{1}{N}-1}\mathcal{I}_{-\frac{1}{N}}} \right) \nonumber \\
                &&=\frac{1}{2}\frac{1}{N A^{2}}\sum_{k=2}^{N-2}\frac{w_k w_{N-k}}{w_{N-1}}\left(\frac{1}{\mathcal{I}_{\frac{k}{N}-1}\mathcal{I}_{-\frac{k}{N}}}-\frac{1}{\mathcal{I}_{\frac{1}{N}-1}\mathcal{I}_{-\frac{1}{N}}} \right) .
             \label{eqn:p1-1}
            \end{eqnarray}
            We conclude that
            \begin{eqnarray} \label{eqn:keff}
                \epsilon_1 \epsilon_2 k_{\rm eff} =\left(-1+\frac{1}{\mathcal{I}_{-\frac{1}{N}}\mathcal{I}_{\frac{1}{N}-1}}\frac{1+q}{1-q}\right) w_1 -\frac{1}{2N}\sum_{k=2}^{N-2}\frac{w_k w_{N-k}}{w_{N-1}}\frac{1+q}{1-q}\left(\frac{1}{\mathcal{I}_{\frac{1}{N}-1}\mathcal{I}_{-\frac{1}{N}}}-\frac{1}{\mathcal{I}_{\frac{k}{N}-1}\mathcal{I}_{-\frac{k}{N}}} \right).
            \end{eqnarray}

        \subsection{Full coupling constants} Comparing (\ref{eqn:keff}) with (\ref{eqn:genF}), we see that the coupling constantans satisfy the differential equations
        \begin{equation}
            2\pi \ii q  \frac{ \dd \tau^{(l)}}{\dd q} =\frac{1+q}{1-q}\frac{1}{ \mathcal{I}_{-\frac{l}{N}}(\kappa)\mathcal{I}_{\frac{l}{N}-1}(\kappa)}
        \end{equation}
        and the boundary conditions follow from the perturbative part
        \begin{equation}
            2\pi \ii \tau^{(l)}=2\pi \ii \tau^{(l)}_{\mathrm{pert}}+\mathcal{O}(q),\quad q\to 0.
        \end{equation}
        Naturally, the solutions are given by
        \begin{equation} \label{eqn:tau-k}
            2\pi \ii \tau^{(l)}=-\frac{\Gamma\left(\frac{l}{N}\right)^2}{\Gamma\left(\frac{2l}{N}\right)}\frac{{{}_2F_1\left(\frac{l}{N},\frac{l}{N},\frac{2l}{N},1-q\right)}}{{}_2F_1\left(\frac{l}{N},\frac{l}{N},1,q\right)}+\pi\left(\cot\left(\frac{\pi l}{N}\right) +\ii\right).
        \end{equation}
        \par It will be convenient to write the coupling constants also as functions of a new variable
        \begin{eqnarray}
            t=\frac{4}{\kappa^2}=\frac{4}{q+2+\frac{1}{q}}.
        \end{eqnarray}
        Acting exactly in the same way as we did in \cite{BykovSysoeva}, we find that (\ref{eqn:tau-k}) can be rewritten as
        \begin{equation}
            2\pi \ii \tau^{(l)}=-\frac{\Gamma\left(\frac{l}{N}\right)^2}{\Gamma\left(\frac{2l}{N}\right)}2^{\frac{l}{N}} \frac{{}_2F_1\left(\frac{l}{2N},\frac{N+l}{2N},\frac{N+2l}{2N};1-t\right)}{{}_2F_1\left(\frac{l}{2N},\frac{N+l}{2N},1;t\right)} + \pi\left(\cot\left(\frac{\pi l}{N}\right) +\ii\right).
        \end{equation}
        \par This result, up to notation, agrees with the $N=2,3,4$ cases considered in \cite{Billoetal} and the $N=4,6$ cases in \cite{Lerda}, as well as with the first orders presented there for $N=5,7$. Moreover, it seems to be in line with the conjecture made in \cite{Lerda} for the general case, but simpler.
        \\ \textbf{Remark.} In the case of $N=4$, one of two independent coupling constants coincides with the unique coupling appearing in $N=2$ theory, as was shown in \cite{Billoetal}. For $N=6$, a similar observation was made in \cite{Lerda}, where, among three independent constants, appeared the unique constants of $N=2$ and $N=3$ theories. From (\ref{eqn:tau-k}) we immediately see that this is no coincidence, and such repetitions occur whenever a fraction $l/N$ can be reduced.

    \section{\texorpdfstring{$S$}{S}- and \texorpdfstring{$T$}{T}-transformations and modular properties} \label{sec:modularproperties}
        Let us now discuss the modular properties of the coupling constants.
        \par Recall that in \cite{Lerda} the decomposition of the coupling matrix was designed so that each coupling constant transforms independently under the $S$-duality group. Our first goal is to show that (\ref{eqn:tau-decomose}) satisfies this property for all $N$.    
        \par 
        The first question we should answer is how the $S$ group action can be expressed in terms of the symmetric variables. 
        \par The $S$ and $T$ duality transformations can be understood as linear maps acting on $2N$-dimensional vectors
        \begin{equation}
            \begin{pmatrix}
                a_1 \\
                \vdots \\
                a_N \\
                \hat{a}_1^D \\
                \vdots \\
                \hat{a}_N^D
            \end{pmatrix}
        \end{equation}
        and preserving the symplectic form
        \begin{equation}
            \Omega=\sum_{u=1}^N a_u \wedge \hat{a}_u^D.
        \end{equation}
        To rewrite the latter condition in terms of the symmetric variables we define
        \begin{equation}
            \Omega':=\sum_{l=1}^{N-1} v_l \wedge v_{N-l}^D= \frac{1}{N}\sum_{u,v=1}^{N} a_v \wedge a_u^D \sum_{l=1}^{N-1}\ee^{\frac{2\pi \ii l (u-v)}{N}}.
        \end{equation}
        Now note that
        \begin{equation}
            \sum_{l=1}^{N-1}\ee^{\frac{2\pi \ii l (u-v)}{N}}=\sum_{l=1}^{N}\ee^{\frac{2\pi \ii l (u-v)}{N}}-1=N\delta_{u,v}-1.
        \end{equation}
        Taking into account the condition $\sum_{v=1}^N a_v=0$, we conclude that $\Omega'=\Omega$.
        Therefore, the $S$ and $T$ dualities should be given by linear maps acting on $2N-2$ dimensional vectors
        \begin{equation}
            \begin{pmatrix}
                \delta{v}_{N-1}^D  \\
                v_1 \\
                \delta {v}_{N-2}^D \\
                v_{2} \\
                \vdots \\
                \delta {v}_{1}^D \\
                 v_{N-1} \\
            \end{pmatrix}
        \end{equation}
        and preserving the symplectic form $\Omega'$, which, up to irrelevant constant factors, in this basis is presented by the matrix
        \begin{equation}
            \begin{pmatrix}
               \sigma_y &0 &0&\cdots &0 & 0 \\
                0  &  \sigma_y & 0 & \cdots & 0& 0 \\
                \vdots & \vdots & \ddots & \ddots & \vdots & \vdots \\
                \vdots & \vdots & \ddots & \ddots & 0 &0 \\
                0 & 0&  \ldots & \ldots & \sigma_y & 0 \\
                0 & 0 & \ldots & \ldots &0 &\sigma_y,
            \end{pmatrix}
        \end{equation}
        where $\sigma_y$ is the Pauli matrix acting on the two-dimensional spaces of the vectors
        \begin{equation}
            \begin{pmatrix}
                \delta v_{N-l}^D\\
                 v_l
            \end{pmatrix},
        \end{equation}
        $l=1,\ldots, N-1$. Here we took into account that, in general, the dual variables can have nonzero values, which have to be treated separately from the variations $\delta v_{l}^D$. We have seen that in the special vacuum this is the case for $l=1$. 
        \par It is natural to expect that $S$- and $T$-dualities commute with the residual Weyl symmetry. Since by (\ref{eqn:vtrans}) both components of a vector
        \begin{equation}
            \begin{pmatrix}
                \delta v_{N-l}^D\\
                 v_l
            \end{pmatrix},
        \end{equation}
        transform under the residual Weyl symmetry with the same eigenvalue, we conclude that $S$- and $T$-dualities can act only by block-diagonal matrices of the form
        \begin{equation}\label{eqn:gammaST}
            \begin{pmatrix}
               \gamma^{(1)} &0 &0&\cdots &0 & 0 \\
                0  &  \gamma^{(2)} & 0 & \cdots & 0& 0 \\
                \vdots & \vdots & \ddots & \ddots & \vdots & \vdots \\
                \vdots & \vdots & \ddots & \ddots & 0 &0 \\
                0 & 0&  \ldots & \ldots & \gamma^{(N-2)} & 0 \\
                0 & 0 & \ldots & \ldots &0 &\gamma^{(N-1)}
            \end{pmatrix}.
        \end{equation}
        Here for $l=1,\ldots,N-1$, the two-by-two matrix $\gamma^{(l)}$ should preserve the standard symplectic form $\sigma_y$, which is equivalent to $\gamma^{(l)}\in SL(2)$.
        \par We now see that the main advantage of introducing the coefficients $\tau^{(l)}$ is, actually, the drastic simplification of the $S$ group action it brings.
        \par From the block diagonal form (\ref{eqn:gammaST}), inherent to any duality respecting residual Weyl symmetry, follows that for each $l$ the variables $v_{l}$ and $\delta v_{N-l}^D$ transform via themselves by a matrix $\gamma^{(l)}\in SL(2)$. But this means that $\gamma^{(l)}$ acts on the coupling constant
        \begin{equation}
            \tau^{(l)}=\frac{1}{2\pi\ii}\frac{\partial^2 \mathcal{F}}{\partial v_l\partial v_{N-l}}=\frac{\partial v_{N-l}^{D}}{\partial v_l}
        \end{equation}
        by the corresponding fractional linear transformation.
        In other words, the constants $\tau^{(l)}$ are exactly the $\lfloor\frac{N}{2}\rfloor$ constants on which the $S$ group acts diagonally, whose existence was predicted in \cite{Lerda}.
        \par Let us now find the form of the matrices $\gamma^{(l)}$ using the $S$ and $T$ transformations from \cite{Lerda,ArgyresBuchel}. 
        The $S$ transform is given by\footnote{To see this, one has to take into account that in \cite{Lerda} the periods along the contours $\alpha_u$ and $\beta_u$ are respectively $a_u$ and $\hat{a}_{u}^D-\hat{a}_{u-1}^D$ }
        \begin{equation}
            S: \ a_u\to \hat{a}_{u}^D-\hat{a}_{u-1}^D,\  \hat{a}_{u}^D-\hat{a}_{u-1}^D\to {a}_{u-1}.
        \end{equation}
        Let us separate the special vacuum values $a_u^{(0)}$, $a_u^{D(0)}$ from the variations $\delta a_u^{(0)}$, $\delta a_u^{D(0)}$. Using (\ref{eqn:au0}) and (\ref{eqn:auD0}) we see that action on the former is
        \begin{equation}
           S: A\to-2\pi\ii \tau^{(1)}A\left (1-\ee^{-\frac{2\pi \ii}{N}}\right), \quad -2\pi\ii\tau^{(1)}A\left (1-\ee^{-\frac{2\pi \ii}{N}}\right)\to A\ee^{-\frac{2\pi \ii }{N}}.
        \end{equation}
        This can be presented in a clearer form
        \begin{equation}
            S:  A\to 4\pi A \tau^{(1)} \sin\left(\frac{\pi}{N}\right)\ee^{-\frac{\pi \ii}{N}},\quad \tau^{(1)} \to 
            -\frac{1}{4\sin^2\left(\frac{\pi}{N}\right)\tau^{(1)}}.
        \end{equation}
        \par The transformation for the coupling constant has the expected form. We shall discuss it later together with the rest of the coupling constants.
        \par The transformation for the parameter $A$ implies that the $S$ duality is not really a symmetry of the special vacuum, but instead transforms one special vacuum into another. Note that $S^2$ acts as
        \begin{equation}
            S^2:\ A\to A \ee^{-\frac{2\pi \ii u}{N}},\ \tau^{(1)} \to \tau^{(1)},
        \end{equation}
        \textit{ i.e.} it realises the residual Weyl symmetry.
        \par
        For the variations, by (\ref{eqn:a-to-v}) and (\ref{eqn:vdviaargcyc}) we have
        \begin{equation}
           S: \  v_l\to \left(1-\ee^{\frac{2\pi\ii l}{N}}\right)\delta {v}_{N-l}^D,\quad \delta v_{N-l}^D\to \frac{1}{(1-\ee^{\frac{2\pi \ii l}{N}})} \ee^{\frac{2\pi \ii l}{N}}v_l^.
        \end{equation}
        This leads to
        \begin{equation}\label{eqn:gammaS-l}
            \gamma_S^{(l)}=
            \begin{pmatrix}
                0 & \frac{1}{-2\ii \sin\left(\frac{\pi l}{N}\right)}\ee^{\frac{\pi\ii l}{N}}\\
                -2\ii \sin\left(\frac{\pi l}{N} \right) \ee^{\frac{\pi\ii l}{N}}& 0
            \end{pmatrix}
            \sim 
            \begin{pmatrix}
                0 & -\frac{1}{2\sin\left(\frac{\pi l}{N}\right)}\\ 2\sin\left(\frac{\pi l}{N}\right) & 0
            \end{pmatrix},
        \end{equation}
        where $\sim$ denotes the projective equivalence in $PSL(2)$. Therefore, on the coupling constants, $S$ acts as
        \begin{equation}\label{eqn:S-transform}
            S:\ \tau^{(l)}\to -\frac{1}{4\sin\left(\frac{\pi l}{N}\right)^2  \tau^{(l)}}. 
        \end{equation}
        Note that this is consistent with already found transformation law for $l=1$.
        \par 
        An analogous analysis for the $T$-transformation
        \begin{equation}
            T: \ a_u\to a_u,  \hat{a}_{u}^D-\hat{a}_{u-1}^D\to \hat{a}_{u}^D-\hat{a}_{u-1}^D+a_u-a_{u-1},
        \end{equation}
        leads to
        \begin{equation}\label{eqn:T-transform}
            T:\ A\to A, \quad\tau^{(l)}\to \tau^{(l)}+1.
        \end{equation}
        \par We conclude that, as it was predicted in \cite{Lerda}, the coupling matrix in general decomposes into a sum of $\lfloor\frac{N}{2}\rfloor$ matrices, transforming independently under the $S$-duality group. The transformation laws we have found are exactly the same as in \cite{Lerda}. In particular, we have found a simple explanation for appearance of $\sin\left(\frac{\pi l}{N}\right)$ in the `spectrum' of S. 
        \par The generators $S$ and $T$ act on the ultra-violet (bare) coupling constant as (\cite{Lerda})
        \begin{equation}
            S:\ q\to q^{-1},\ T:\ q\to q.
        \end{equation}
        So, the functions $q_{UV}^{(l)}$ inverse to $\tau^{(l)}$ should be invariant under action of the group generated by
        \begin{equation}
            T,\ STS.
        \end{equation}
        By rescaling $\tau^{(l)}$ in the same manner as in \cite{BykovSysoeva}, we recognise in the matrices $\tilde{\gamma}_{T}^{(l)}$ and $\tilde{\gamma}_{S}^{(l)}$ generators of the triangle group $\Gamma_{(\frac{N}{N-2l},\infty,\infty)}$, an an index two subgroup of the Hecke group $\Gamma_{(\frac{2N}{N-2l},2,\infty)}$ used in \cite{Lerda}. Thus, $q$ is a modular function (and, in fact, the Hauptmodul, see \cite{BykovSysoeva}) for the group $\Gamma_{(\frac{N}{N-2l},\infty,\infty)}$, and $t$ is a modular function for both $\Gamma_{(\frac{N}{N-2l},\infty,\infty)}$ and $\Gamma_{(\frac{2N}{N-2l},2,\infty)}$ (and a Hauptmodule for the latter). 
        \par 
        We recall that it is expected that the Seiberg-Witten curves would admit presentations with modular forms of the coupling constants as coefficients. Let us examine whether the form (\ref{eqn:structure}) possesses this property.
        \par The coefficient $\kappa=\sqrt{q}+\frac{1}{\sqrt{q}}$ is expressed in terms of the $\Gamma_{(\frac{N}{N-2l},\infty,\infty)}$-modular function $q$, but contains a square root, which may in principle introduce an additional sign.
        \par The coefficients $p_k$ in (\ref{eqn:pkleadingmassless}) are expressed in terms of the functions $\mathcal{I}_{l\alpha}$, which are closely related to the weight-two $\Gamma_{(\frac{N}{N-2l},\infty,\infty)}$-modular form $f_2^{(l)}$ introduced in Appendix \ref{app:modularprop}. In fact, multiplying numerator and denominator of (\ref{eqn:pkleadingmassless}) by $(1-t)^{\frac{k+1}{2N}}$ and using Appendix \ref{app:modularprop}, we get 
        \begin{equation}
            p_k^{(0)}=-\frac{w_k}{A^{k+1}} \sqrt{\frac{t-1}{f_2^{(k)}\left(-f_2^{(1)}\right)^k}} +\mathcal{O}(\delta^2).
        \end{equation}
        We note that the factor $\sqrt{1-t}=\frac{1-q}{1+q}$ is a $\Gamma_{(\frac{N}{N-2l},\infty,\infty)}$-modular function, while the factor $(1-t)^{\frac{k}{2N}}$, which has a nontrivial monodromy at $t=1$ and thus would necessarily affect the modular properties\footnote{The relation between the modular properties and monodromy group, for the case of the triangular groups, is extensively discussed in \cite{Doran}. See also \cite{BykovSysoeva} for application to our setup.}, cancelled. Nevertheless, unexpected square roots of the modular forms still appear. \par Finally, the next order correction $p_1^{(1)}$ can be expressed in terms of the modular forms $f_2^{(l)}$ as
        \begin{equation}
            p_1^{(1)}=\frac{1}{2}\frac{1}{N A^{2}}\sum_{k=2}^{N-2}\frac{w_k w_{N-k}}{w_{N-1}}\left(-\frac{\sqrt{1-t}}{f_2^{(k)}}+\frac{\sqrt{1-t}}{f_2^{(1)}} \right). 
        \end{equation}
        \par We conclude that (\ref{eqn:structure}) is almost, but not quite, in the desired form, because the coefficients contain square roots of the modular objects. One may hope, however, that it is possible to eliminate the square roots by a change of variables.
        \\ \textbf{Remark.} The $S$-duality group in general acts nontrivially on the parameter $A$ that determines the special vacuum even if we restrict our attention to the subgroup  $\Gamma_{(\frac{N}{N-2l},\infty,\infty)}$. However, as long as we work with the massless case (or with the asymptotic regime \(A\to\infty\)) the coupling constants do not depend on $A$, so this issue can be ignored.

    \section{Higher order corrections to the Seiberg-Witten curve}  \label{sec:highord}
        In the previous sections, we found the  coefficients $p_i$ of the function $y(z)$ up to second order in $\delta$. In this section, we look for a convenient form of all higher orders, allowing a more general right-hand side, and see how higher-order corrections affect the periods. The main purpose of this computation is to automate the introduction of the mass corrections in the next Section. However, it can be used for other purposes, such as finding the derivatives of the coupling constants with respect to $a_u$.  
        \par 
        We consider an equation of the form
        \begin{equation}\label{eqn:SW-allord-first}
            y^{-N}\left(y^N-U^N+U^NG(y)\right)=(y^{(0)}_u)^{-N}\left(y^{(0)N}_u-U^N\right),
        \end{equation}
        where
        \begin{equation}
            G(y)=\frac{1}{U^N}\sum_{k=1}^{\infty}h_k y^{N-k}.
        \end{equation}
        We can rewrite (\ref{eqn:SW-allord-first}) as  
        \begin{equation}  \label{eqn:SW-allord}
            y^{-N}\left(1-\frac{1}{U^N}\sum_{k=1}^{\infty}h_k y^{N-k}\right)=(y^{(0)}_u)^{-N}.
        \end{equation}
        If we set
        \begin{equation}
            h_k=p_{k-1}(-1)^k,\ k=1,\ldots,N-1
        \end{equation}
        and $h_k=0$ for all other values of $k$, we recover the equation $\omega(y(z))=z$ with $\omega$ defined by (\ref{eqn:structure}) and with $y_u^{(0)}$ given by (\ref{eqn:y-cut}). However, it is more convenient to define $U$ by its leading order (\ref{eqn:U-solvec}), and absorb all its corrections into the coefficient $h_N$ by setting $h_{N}=(-1)^N p_{N-1}+U^N$.
        \par 
        We now allow all the coefficients $h_k$ to be nonzero, but small, and treat equation (\ref{eqn:SW-allord}) perturbatively. We are looking for a solution of the form
        \begin{equation}\label{eqn:yu-h-lead}
            y_u=y_u^{(0)}+O(h_1,h_2,\ldots),
        \end{equation}
        which fixes a branch of the multivalued function $y(z)$. 
        \par
        From (\ref{eqn:SW-allord}) follows that for an arbitrary degree $r$ 
        \begin{equation}
            (y_u)^r=(y^{(0)}_u)^r\left(1-\frac{1}{U^N}\sum_{k=1}^{\infty}h_k y_u^{N-k}\right)^{\frac{r}{N}}.
        \end{equation}
       Here we took (\ref{eqn:yu-h-lead}) into account  to choose the branch of the rational power (which is assumed to be the standard one). Expanding the parentheses, we get
        \begin{equation}
         \label{eqn:yr-req}   y^r_u=(y^{(0)}_u)^r\sum_{n=0}^{\infty}\frac{\left(-\frac{r}{N}\right)_n}{n!U^{nN}}\sum_{k_1,\ldots,k_n}h_{k_1}\cdots h_{k_n} y_u^{n N-k_1-\ldots-k_n},
        \end{equation}
        where $\left(-\frac{r}{N}\right)_n$ is the Pochhammer symbol.
        This equation can be solved iteratively. It is easy to see that its solution has the form
        \begin{equation}
            \label{eqn:yr-ansatz}y^r_u=\sum_{n=0}^{\infty}\sum_{k_1,\ldots,k_n}\frac{h_{k_1}\ldots h_{k_n}}{n! U^{n N}}\left(y^{(0)}_u\right)^{r+n N-k_1-\ldots-k_n}B^n_{r,N}(k_1,\ldots,k_n),
        \end{equation}
        where $B^n_{r,N}(k_1,\ldots,k_n)$ are some combinatorial coefficients. Remarkably, the coefficients $B^n_{r,N}(k_1,\ldots,k_n)$ depend only on the sum $k=k_1+\ldots+k_n$ and can be written as
        \begin{equation}\label{eqn:surprise}
           B^n_{r,N}(k)=B_{r,N}^{n,k}=\frac{r}{N^n} \prod_{t=1}^{n-1}(r+t N-k)=\frac{r}{N}\left(\frac{r-k}{N}+1\right)_{n-1}(-1)^n.
        \end{equation}
        For $n=0$, the Pochhammer symbol should be understood as
        \begin{eqnarray}
            (x)_{-1}=\frac{1}{x-1},
        \end{eqnarray}
        so $B^{0,0}_{r,N}=1$ (clearly, for $n=0$, $k_1+\ldots+k_n=0$, so we do not need to consider other values of $k$), which ensures that (\ref{eqn:yr-ansatz}) is compatible with (\ref{eqn:yu-h-lead}).
        This can be proven by induction in $n$, substituting (\ref{eqn:yr-ansatz}-\ref{eqn:surprise}) into (\ref{eqn:yr-req}), and summing the coefficients in front of $h_{k_1}\cdots h_{k_n}$.
        \par In (\ref{eqn:yr-ansatz}), $y_u$ is not defined iteratively, so we no longer need the arbitrary parameter $r$. We set $r=1$ and define
        \begin{equation}
            B^{n,k}_N= B^{n,k}_{N,1}=\frac{1}{N}\left(\frac{1-k}{N}+1\right)_{n-1}(-1)^n=-\frac{1}{N}\left(1+\frac{k-1}{N}-n\right)_{n-1}.
        \end{equation}
        Finally, we can formally rewrite
        \begin{equation}
            y_u=y_u^{(0)}\sum_{n_1=0}^{\infty}\sum_{n_2=0}^{\infty}\cdots\Bigg(\prod_{k=1}^{\infty}\frac{\left(\frac{h_{k}}{U^N}\left(y_u^{(0)}\right)^{N-k}\right)^{n_k}}{n_k! }B^{\sum_k n_k,\sum_k k n_k}_{N}\Bigg).
        \end{equation}
        Here we formally have a sum over infinitely many variables $n_1,n_2,\ldots$ but under our assumptions on $h_k$, only finitely many of them are nonzero at each order of the perturbation theory.
        \par We note that
        \begin{equation}
            \underset{\substack{\sum_k n_k=n\\
            \sum_k k n_k=K}}{\sum_{n_1=0}^{\infty}\sum_{n_2=0}^{\infty}\cdots}\Bigg(\prod_{k=1}^{\infty}\frac{\left(\frac{h_{k}}{U^N}\left(y_u^{(0)}\right)^{N-k}\right)^{n_k}}{n_k! }\Bigg)=\left[\frac{1}{n!}G(s)^n\right]_{s^{nN-K}}\left(y_u^{(0)}\right)^{n N-K},
        \end{equation}
        where we introduced the following notation for coefficient extraction from a power series:
        \begin{equation}
            \left[\sum_{k}a_k s^{k}\right]_{s^k_0}:=a_{k_0}.
        \end{equation}
        Then
        \begin{gather}
            y_u=-\sum_{K=-\infty}^{+\infty}\frac{1}{N}\sum_{n=0}^{\infty}\left[\frac{G(s)^n}{n!}\right]_{s^{nN-K}}\left(1+\frac{K-1}{N}-n\right)_{n-1}\left(y_u^{(0)}\right)^{nN-K+1}= \nonumber \\
            -\sum_{K=-\infty}^{+\infty}\frac{1}{N}\sum_{n=0}^{\infty}\left[\frac{G(s)^n}{n!}\right]_{s^{-K-1}}\left(1+\frac{K}{N}\right)_{n-1}\left(y_u^{(0)}\right)^{-K}= \nonumber
            \\
            -\sum_{K\neq 0}  \left[\frac{(1-G(s))^{-\frac{K}{N}}}{K}\right]_{s^{-K-1}}\left(y_u^{(0)}\right)^{-K}+\left[\frac{\ln(1-G(s))}{N}\right]_{s^{-1}}. \label{eqn:deformedcurve}
        \end{gather}
        The equation above is more convenient for both symbolic computations in a computer algebra system and further analysis than the original formula (\ref{eqn:yr-ansatz}).
        % \par
        % Moreover, we shall see that the logarithmic term is not relevant for what follows. In fact, one can verify that in the cases of interest it has to vanish due to the restrictions imposed on $H(y)$ by the asymptotic behaviour of $\omega$. 
        \par 
        We can now write the periods (\ref{eqn:periodsdz-v}) for the curve given by (\ref{eqn:deformedcurve}). 
        \begin{equation}
            -\delta_{l,N-1}-\frac{v_{l}}{A\sqrt{N}}=-\sum_{f=-\infty}^{\infty} \left[\frac{(1-G(s))^{-\frac{l-fN}{N}}}{l-fN}\right]_{s^{fN-l-1}}U^{fN-l}\mathcal{I}_{\frac{l}{N}-f},
        \end{equation}
        or 
        \begin{equation}\label{eqn:vl-G}
            -\delta_{l,N-1}-\frac{v_{l}}{A\sqrt{N}}=-{\rm Res}_{s=0}\sum_{l'\equiv l \, \mod\, N} s^{l'}\frac{(1-G(s))^{-\frac{l'}{N}}}{l'}
            U^{-l'}\mathcal{I}_{\frac{l'}{N}}.
        \end{equation}
        
        Here we took into account that only the terms of the form $\left(y_u^{(0)}\right)^{fN-l}$ contribute to the equation for $v_l$. In particular, the constant logarithmic term in (\ref{eqn:deformedcurve}) does not affect $v_l$. 
        \par 
        Using the techniques developed in Appendix \ref{app:universalf}, we can express the coefficients of these equations in terms of the modular and quasi-modular forms of the groups $\Gamma_{(\frac{2N}{N-2l},2,\infty)}$ and $\Gamma_{(\frac{N}{N-2l},\infty,\infty)}$, which suggests that all reasonable deformations of the coupling constants enjoy some form of modularity. Though, it is difficult to make this vague statement more precise at this level of generality.
        \par
        For the sake of completeness, we also provide an explicit formula in terms of $h_k$;
        \begin{equation}
            -\delta_{l,N}-\frac{v_{l-1}}{A\sqrt{N}}=\sum_{n=0}^{\infty}\sum_{\substack{k_1,\ldots,k_n \\ k_1+\cdots+k_n \equiv l \ \mod N}}B_{N}^{n,k_1+\cdots+k_n}\frac{h_{k_1}\cdots h_{k_n}}{n!}\mathcal{I}_{\frac{k_{1}+\ldots+k_n-1}{N}-n}(\kappa)U^{1-k_1-\ldots-k_n}.
        \end{equation}

    \section{Mass corrections to the coupling constants} \label{sec:masscorrections}
    
        \par In this section, we study how nonzero masses of the multiplets affect the coupling constants and their modular properties.
        \par Although we will see how far we can proceed with the general mass distribution, the interesting results will be achieved for the mass distribution respecting the $\mathbb{Z}_n$ symmetry.
        \par For the $\mathbb{Z}_n$ symmetric masses we introduce the notation
        \begin{equation} \label{eqn:specialmasses}
            \begin{cases} 
                m_f=m \ \ee^{\frac{2\pi \ii f}{N}} , \quad f=1,\ldots,N , \\
                 m_f=m' \ \ee^{\frac{2\pi \ii f}{N}} , \quad f=N+1,\ldots,2N
            \end{cases}
        \end{equation}
        and we denote by $\bm m$ the $2N$-tuple of masses (\ref{eqn:specialmasses}).
        \par This mass spectrum coincides with the massive special vacuum considered in \cite{Lerda}.
        
        \subsection{Perturbative part of the prepotential deformed by masses}
            The classical contribution to the prepotential does not change upon adding masses; however, the one-loop part does.
            \begin{equation}
                \mathcal{F}_{{\rm 1loop},\bm m}=\frac{1}{2}\sum_{\substack{u,v=1 \\  u \neq v}}^N (a_u-a_v)^2 \ln \frac{a_u-a_v}{\Lambda} - N \sum_{u=1}^N \sum_{f=1}^{2N} (a_u+m_f)^2 \ln \frac{a_u+m_f}{\Lambda}
            \end{equation}
            Then the second derivatives of the prepotential in the presence of nonzero masses are
            \begin{eqnarray}
                \hat{\partial}_u\hat{\partial}_v \mathcal{F}_{{\rm 1loop},\bm m}&=& \delta_{u,v} \left(\sum_{\substack{k=1 \\ k \neq u}}^N \ln \left(-\frac{(a_{u}-a_k)^2}{\Lambda^2}  \right)-\frac{1}{2}\sum_{f=1}^{2N}\ln \left( \frac{(a_u+m_f)^2}{\Lambda^2} \right) \right) \nonumber \\&+&(1-\delta_{u,v})\left(-\ln\left(-\frac{(a_{u}-a_v)^2}{\Lambda^2}\right)\right) + \rm{const} .
            \end{eqnarray}
            \par For general masses the dependence on the scale $\Lambda$ does not disappear. With $m_f=0$ for all $f$, this dependence reduces to a contribution to the final constant term, which disappears in the next step and does not affect the coupling matrix. Let us now assume the special choice of nonzero masses (\ref{eqn:specialmasses}). Then
            \begin{equation}
                \frac{1}{2}\sum_{f=1}^{2N}\ln \left(\frac{(a_u+m_f)^2}{\Lambda^2} \right)=N\ln \left(\frac{a_u^2}{\Lambda^2} \right)+\ln\left(\prod_{f=1}^{2N}\left(1+\frac{m_f}{a_u}\right) \right)=N\ln \left(\frac{a_u^2}{\Lambda^2} \right)+\ln\left(1+\frac{T}{A^N}+\frac{T'}{A^{2N}} \right),
            \end{equation}
            where $T=m^N+m'^N$, $T'=m^N m'^N$.
            \par Then the variation of the prepotential is
            \begin{eqnarray}
                &&\delta\mathcal{F}_{{\rm 1loop},\bm m}=\delta\mathcal{F}_{\rm 1loop}-\frac{1}{2}\sum_{u,v=1}^N \delta a_u \delta a_v \delta_{u,v} \ln\left(1+\frac{T}{A^N}+\frac{T'}{A^{2N}} \right).
            \end{eqnarray}
            Proceeding as in the massless case, we obtain
            \begin{eqnarray}
                &&\delta\mathcal{F}_{{\rm 1loop},\bm m}=  \\
                &&\frac{1}{2}\sum_{l=1}^{N-1}v_l v_{N-l} \Big( 2 \ln 2 N+ \pi \ii (N-1)  -  \sum_{k=1}^{N-1} \cos\left(\frac{2 \pi  k l}{N}\right)\ln\left(\sin^2\frac{\pi k}{N}\right)-\ln\left(1+\frac{T}{A^N}+\frac{T'}{A^{2N}} \right) \Big). \nonumber 
            \end{eqnarray}
            Hence, the perturbative contribution to the coupling constant is
            \begin{equation}
                2\pi\ii \tau^{(l)}_{\mathrm{pert}, {\bm m}}=\ln(q_0)- 2 \ln 2 N- \pi \ii (N-1)  +  \sum_{k=1}^{N-1} \cos\left(\frac{2 \pi  k l}{N}\right)\ln\left(\sin^2\frac{\pi k}{N}\right)-\ln\left(1+\frac{T}{A^N}+\frac{T'}{A^{2N}} \right). 
            \end{equation}
            Rewriting it in terms of $q=(-1)^N q_0$ and choosing the branch of the logarithm such that 
            \begin{equation}
                \ln(q)=\ln(q_0)-\pi \ii N
            \end{equation}
            we arrive at 
            \begin{equation}\label{eqn:tau-pert-m}
                2\pi\ii \tau^{(l)}_{\mathrm{pert}, {\bm m}}=\ln(q)- 2 \ln 2 N +  \sum_{k=1}^{N-1} \cos\left(\frac{2 \pi  k l}{N}\right)\ln\left(\sin^2\frac{\pi k}{N}\right)+\pi \ii-\ln\left(1+\frac{T}{A^N}+\frac{T'}{A^{2N}} \right). 
            \end{equation}

        \subsection{Seiberg-Witten curve deformed by masses}
            Let us now apply the approach of Section \ref{sec:highord} to find the mass corrections. We start by assuming that the masses of the hypermultiplets are nonzero, $\frac{m_i}{A}\sim \varepsilon \ll 1$ for each flavour index $i$, and are otherwise arbitrary. Note that the parameter $\delta$ introduced in Section \ref{sec:pertutbation} and $\varepsilon$ are independent. 
            \par Define
            \begin{equation}
                \tilde{Q}(y)=\prod_{j=1}^{2N}(y-m_j)=y^{2N}-\tilde{T}_0 y^{2N-1}+\tilde{T}_1 y^{2N-2}+\cdots-\tilde{T}_{N-1},
            \end{equation}
            where 
            \begin{equation}
                \tilde{T}_l=\sum_{i_0<\ldots<i_l}m_{i_0}\cdots m_{i_l}.
            \end{equation}
            Here we adopt the notation of \cite{BykovSysoeva}.
            \par Define also
            \begin{equation} \label{eqn:Q}
                Q(y)=\sqrt{\tilde{Q}(y)}.
            \end{equation}
            \par Note that in the case $m_i=m_{i+N}$, $i=1,\ldots,N$ (\textit{e.g.} for $N$ pairs of fundamental and anti-fundamental multiplets of equal masses), $Q(y)$ is a polynomial. Such a definition of $Q(y)$ is also in agreement with \cite{BykovSysoeva}, since in the present case $\epsilon_1=\epsilon_2=0$.
            \par In the case of a general mass distribution, $Q(y)$ defined by (\ref{eqn:Q}) is a multivalued function. Still, for $m_i\ll A$ we can always choose its branch so that the cuts are located near the origin and far from the cuts of $\omega$. For the sake of definiteness, we assume that $Q(y)= y^N +\mathcal{O}(y^{-1})$ for $y\to \infty$. It is easy to see that under such assumptions the analysis of \cite{BykovSysoeva}, where $Q$ was assumed to be a polynomial, can be applied without modification. So, the Seiberg-Witten curve, which we have already described by the function $y(z)$, takes the form
            \begin{equation}\label{eqn:SW-mass}
                z+\frac{1}{z}=\kappa A^N \frac{P_N(y(z))}{Q(A y(z))}.
            \end{equation}
            \par The masses also affect the asymptotic behaviour of $\omega$, and thus the connection of $p_1$ with $k_{\mathrm{eff}}$. For $y\to \infty$, we have
            \begin{equation}
                p_1=\frac{q \tilde{T}_1+(w_1+k_{\rm eff} \epsilon_1 \epsilon_2)(1-q)}{A^2(1+q)}.
            \end{equation}
            But in the special vacuum $\tilde{T}_1\neq 0$ only for $N=2$, where $\tilde{T}_1=T=m^2+m'^2$. Hence, for $N \geq 3$ the relation between $p_1$ and $k_{\rm eff}$ remains the same, while for $N=2$ it shifts as
            \begin{equation}
                \epsilon_1 \epsilon_2 k_{\rm eff} =\frac{A^2 p_1 (1+q)- q T}{1-q}-w_1.
            \end{equation}
            \par It is easy to see that equation (\ref{eqn:SW-mass}) is equivalent to an equation of the form (\ref{eqn:SW-allord}) with the coefficients $h_k$ defined by
            \begin{gather}
                U^NG(y)=\sum_{k=1}^{\infty} h_k y^{N-k}=y^N A^N \frac{P_N(y)}{Q(A y)}-\left(y^N-U^N\right)= \nonumber
                \\G_1(y)\left(\sum_{k=2}^{N-1}(-1)^k p_{k-1} y^{N-k}+(-1)^{N}\delta p_{N-1}\right)+\left(y^N-U^N\right)(G_1(y)-1),\label{eqn:h-def-ser}
            \end{gather}
            where $\delta p_{N-1}=p_{N-1}-(-1)^N U^N$ is the deviation of $p_{N-1}$ from its massless special vacuum value, and
            \begin{equation}\label{eqn:G1-Def}
                G_1(y)=\frac{A^Ny^N}{Q(Ay)}.   
            \end{equation}
            \par
            In order to apply the perturbative techniques of Section \ref{sec:highord}, we have to show that all $h_k$ carry a small-order parameter, either $\varepsilon$ or $\delta$.
            \par
            From (\ref{eqn:G1-Def}) follows that $G_1(y)=1+\mathcal{O}(\varepsilon)$, so the second term in (\ref{eqn:h-def-ser}) is small and of order $\mathcal{O}(\varepsilon)$. 
            \par 
            The coefficients $p_k$, $k=1,\ldots,N-2$ characterise how much the residual Weyl symmetry of the exact massless special vacuum is broken by the perturbations. In general, they can contain both terms proportional to $\varepsilon$ and $\delta$, but absent in the massless special vacuum, as we saw in (\ref{eqn:pkleadingmassless}). 
            \par
            Finally, from the Fourier analysis we see that the parameter $\delta p_{N-1}$ is determined by the equation (\ref{eqn:vl-G})  with $l=N-1$. Essentially, its role is to compensate the perturbation $\delta w_{N-1} \propto v_{N-1}$, which should be small by definition, caused by the other terms of $G(s)$. Hence, $\delta p_{N-1}$ is small whenever the rest of the terms of $G(s)$ are.
            \par We conclude that all $h_k$ in (\ref{eqn:h-def-ser}) are small and we can safely use the techniques of  Section \ref{sec:highord}. 
     
            %where we took into account $A\gg m_i$.
            We illustrate how it works in case of the special mass spectrum (\ref{eqn:specialmasses}), which corresponds to 
            \begin{equation} \label{eqn:Q-spec}
               Q(x)^2= \tilde{Q}(x)=(x^{N}-m^N)(x^{N}-m'^N)=x^{2N}-T x^N+T'.
            \end{equation}
            \par The special mass spectrum is invariant under the residual Weyl symmetry. Therefore, the symmetry is broken only by $v_l\sim \mathcal{O}(\delta)$.  As a consequence,  $p_l=\mathcal{O}(\delta)$ for $l=1,\ldots,N-2$. Note again the special role of $\delta p_{N-1}$, which, unlike all the others $p_l$, is not associated with the breaking of the residual Weyl symmetry.
            \par It is convenient for us to write (\ref{eqn:vl-G}) in such a way that the power series in terms of $\delta$ is explicit. For that we observe
            \begin{equation}\label{eqn:G-p-V}
                (1-G(s))^{-\frac{l'}{N}}=\sum_{n=0}^{\infty}\sum_{k_1=2}^{N-1}\ldots \sum_{k_n=2}^{N-1} (-1)^{k_1+\ldots+k_n}\frac{p_{k_1-1}\cdots p_{k_n-1}}{n!}s^{nN-k_1-\ldots-k_n}V^{(n)}_{l'}(\delta p_{N-1},s).
            \end{equation}
            Here the sum over $n$ is a sum over the powers of the small parameter $\delta$. 
            \begin{equation}\label{eqn:Vln}
                V^{(n)}_{l'}(\delta p_{N-1},s)=(-1)^{nN}\frac {\partial^{n}}{\partial \delta p_{N-1}^n}V(\delta p_{N-1},s)^{-\frac{l'}{N}},
            \end{equation}
            \begin{equation}\label{eqn:Vdef}
                V(\delta p_{N-1},s)=1-(-1)^{N}\delta p_{N-1} U^{-N}G_1(s)-U^{-N}\left(s^N-U^N\right)(G_1(s)-1).
            \end{equation}
            \par In the special vacuum case, $G_1(s)$, and thus $V^{(n)}_k$, contains only integer powers of $s^N$. As a consequence, the total power of $s$ in (\ref{eqn:G-p-V}) is equal to $k_1+\ldots+k_n$ modulo $N$. This allows to rewrite it as
            \begin{equation}
                A\sqrt{N}\delta_{l,N}+v_l=-A\sqrt{N}\sum_{n=0}^{\infty}\underset{k_1+\ldots+k_n\equiv l+1 \ \mod N}{\sum_{k_1=2}^{N-1}\cdots \sum_{k_n=2}^{N-1}}(-1)^{k_1+\ldots+k_n}\frac{p_{k_1-1}\cdots p_{k_n-1}}{n!}\nu^{(n)}_{k_1+\ldots+k_n}(\delta p_{N-1}),
            \end{equation}
            where
            \begin{equation}\label{eqn:nu-def}
                \nu_k^{(n)}(\delta p_{N-1})=- \sum_{l'\equiv k-1 \, \mod N}{\rm Res}_{s=0}\left(s^{n N-k+l'}U^{-l'} \frac{V_{l'}^{(n)}(\delta p_{N-1})}{l'}\mathcal{I}_{\frac{l'}{N}}\right).
            \end{equation}
            \par For practical use, it is usefull to observe that $V_{l'}^{(n)}$ is a power series consisting of the terms of the form $\varepsilon^{r N}s^{-fN}$, $0\leq f\leq r-n$ (assuming that $\delta p_{N-1}=\mathcal{O}(\varepsilon)$). So, the sum over $l'$ in reality is always bounded, but it is convenient to keep it in this form.  
            By construction,
            \begin{equation}\label{eqn:diflin}
                \nu_k^{(n)}(\delta p_{N-1})=(-1)^{nN}\frac {\partial^{n}}{\partial \delta p_{N-1}^n} \nu_{k-nN}^{(0)}(\delta p_{N-1}).
            \end{equation}
            More explicit formulas for $\nu$ are presented in Appendix \ref{sub-app:nukn}. 
        
            \par Let us now consider the linear order in $\delta$ (and thus in terms of $p_l$, $l=1,\ldots,N-2$). For $l=2,\ldots,N-2$, we have
            \begin{equation}
                v_{l}=A\sqrt{N}(-1)^{l}p_l\nu^{(1)}_{l+1}(\delta p_{N-1})+\mathcal{O}(\delta),
            \end{equation}
            so
            \begin{equation}\label{eqn:pkleading-massive}
                p_{l}=(-1)^{l}\frac{v_l}{A\sqrt{N}\nu_{l+1}^{(1)}(\delta p_{N-1})}+\mathcal{O}(\delta).
            \end{equation}
            For the coefficient $p_1$, we need also the subleading order:
            \begin{equation}
                v_{1}=-A\sqrt{N}p_1\nu^{(1)}_{2}(\delta p_{N-1})-(-1)^N A\sqrt{N}\sum_{l'=2}^{N-2}\frac{p_{l'}p_{N-l'}}{2}\nu^{(2)}_{N+2}(\delta p_{N-1})+\mathcal{O}(\delta^2),
            \end{equation}
            so
            \begin{equation}\label{eqn:p1-masses-first}
                p_1=-\frac{v_1}{A\sqrt{N}\nu_2^{(1)}(\delta p_{N-1})}+\frac{\nu^{(2)}_{N+2}(\delta p_{N-1})}{\nu^{(1)}_{2}(\delta p_{N-1})} \sum_{l'=2}^{N-2}\frac{v_{l'}v_{N-l'}}{2A^2N}\frac{1}{\nu^{(1)}_{l'+1}(\delta p_{N-1})\nu^{(1)}_{N-l'+1}(\delta p_{N-1})}+\mathcal{O}(\delta^3).
            \end{equation}
            \par The special case $l=N-1$ yields an equation determining $\delta p_{N-1}$:
            \begin{equation}
                -v_{N-1}=A\sqrt{N}\left(\nu_{0}^{(0)}(\delta p_{N-1})+1\right)+\mathcal{O}(\delta^2).
            \end{equation}
            This is a nonlinear equation that can be solved order by order. Its  solution can be symbolically written as a formal power series, but we do not find it very useful. It is convenient instead to separate a $v_{N-1}$-dependent part of $\delta p_{N-1}$ from a shift caused by the masses only.  
            \par Let $\overline{\delta p_{N-1}}$ be the solution of
            \begin{equation}\label{eqn:deltapL-bar}
                \nu_{0}^{(0)}(\overline{\delta p_{N-1}})=-1.
            \end{equation}
            We set
            \begin{equation}
                \nu^{(n)}_k=\nu^{(n)}_k(\overline{\delta p_{N-1}}).
            \end{equation}
            Then, using (\ref{eqn:diflin}), we get 
            \begin{equation}
                \delta p_{N-1}=\overline{\delta p_{N-1}}-(-1)^N\frac{v_{N-1}}{A\sqrt{N}\nu^{(1)}_N}+\mathcal{O}(\delta^2).
            \end{equation}
            This allows to rewrite (\ref{eqn:p1-masses-first}) in a more symmetric form
            \begin{gather}\label{eqn:p1-masses-good}
                p_1=-\frac{v_1}{A\sqrt{N}\nu_2^{(1)}}+\frac{\nu^{(2)}_{N+2}}{\nu^{(1)}_{2}}\sum_{l=1}^{N-1}\frac{v_{l}v_{N-l}}{2A^2N}\frac{1}{\nu^{(1)}_{l+1}\nu^{(1)}_{N-l+1}}+\mathcal{O}(\delta^3).%=
                % \\
                % \nonumber\frac{1}{A^2}\left(
                % -\frac{w_1}{A^2 N \nu_2^{(1)}}-\frac{1}{2N}\sum_{l=1}^{N-1}\frac{w_{l}w_{N-l}}{w_{N-1}}\left((-1)^N \frac{\nu^{(2)}_{N+2}}{\nu^{(1)}_{2}} \frac{1}{\nu^{(1)}_{l'+1}\nu^{(1)}_{N-l'+1}}\right)\right)
            \end{gather}
            \par
            Before turning to the coupling constants, let us note that although the coefficients \(\nu_k^{(l)}\) are deformed by the masses, the general structure of (\ref{eqn:pkleadingmassless}) and (\ref{eqn:p1-1}) is preserved. Namely, the leading order of $p_k$ is proportional to $v_k$, and the leading contribution to $p_1$ is a linear combination of $v_1$ and the products $v_lv_{N-l}$. 
            This is a consequence of the fact that the special mass spectrum is residual Weyl symmetry-invariant. 
            \par
            
        \subsection{Differential equations imposed on the deformed coupling constants}\label{ssec:diffeq}
            Combining (\ref{eqn:p1-masses-good}) with (\ref{eqn:p1-keff}) and (\ref{keff-def}), we conclude that the prepotential, as expected, has the form (\ref{eqn:FviaV}) with the deformed couplings $\tau_{{\rm IR},\bm{m}}$, $\tau^{(l)}_{\bm{m}}$ subject to the differential equations
            \begin{equation}\label{eqn:tau-nu}
                2\pi \ii q\frac{\dd \tau_{{\rm IR},\bm{m}}}{\dd q}=\frac{1+q}{1-q}\frac{1}{N\nu_2^{(1)}},
            \end{equation}
            \begin{equation}\label{eqn:tau-l-nu}
                2\pi \ii q\frac{\dd \tau^{(l)}_{\bm{m}}}{\dd q}=- \frac{1+q}{1-q}\frac{\nu^{(2)}_{N+2}}{N\nu^{(1)}_{2}}\frac{1}{\nu^{(1)}_{l+1}\nu^{(1)}_{N-l+1}},
            \end{equation}    
            and the boundary condition for the second equation is determined by the perturbative part:
            \begin{equation} \label{eqn:bound1}
                 2\pi \ii \tau^{(l)}_{\bm{m}}= 2\pi \ii \tau_{{\rm pert}, {\bm m}}^{(l)}+\mathcal{O}(q),\quad q\to 0.
            \end{equation}
            \par In contrast to the conformal massless case, here $\tau_{{\rm IR},\bm{m}}\neq \tau^{(1)}_{\bm{m}}$. Instead, one can verify that (\ref{eqn:tau-nu}-\ref{eqn:tau-l-nu}) are compatible with a differential equation
            \begin{equation} \label{eqn:averagetau}
                \frac{\partial (A\tau_{{\rm IR},\bm{m}})}{\partial A}=\tau^{(1)}_{\bm{m}}
            \end{equation}
            derived in \cite{BykovSysoeva}. Unlike (\ref{eqn:tau-nu}), equation (\ref{eqn:averagetau}) is supplemented by the boundary condition:
            \begin{equation} \label{eqn:bound2}
                \tau_{{\rm IR},\bm{m}}=\tau_{{\rm IR}}+\mathcal{O}(A^{-1})=\tau^{(1)}+\mathcal{O}(A^{-1})=\tau^{(1)}_{\bm{m}}+\mathcal{O}(A^{-1}), \quad A \to \infty .
            \end{equation}
    
             \par Unfortunately, contrary to the massless or large $A$ limits, where the expected modular properties guided the analysis, in the present case we do not have a simple ansatz for the solution, so a more thorough analysis is required. However, the equations (\ref{eqn:tau-nu},\ref{eqn:averagetau}) with the boundary conditions (\ref{eqn:bound1}, \ref{eqn:bound2}) are sufficient for the order-by-order computation of the coupling constants as power series in $q$ and $\delta$, and may also be useful in practice. Examples of such computations for $\tau^{(l)}_{\bm m}$ can be found in Appendix \ref{app:examplestau}. 
        
        \subsection{Observations on the modularity of the deformed couplings}
        \label{ssec:mod-obs}
            To analyse (\ref{eqn:tau-l-nu}-\ref{eqn:bound1}) and derive a closed algebraic (rather than differential) equation for the couplings, it is convenient to express (\ref{eqn:tau-l-nu}) in terms of the modular and quasi-modular forms (see Appendix \ref{app:modularprop}). To this end, we introduce
            \begin{equation}
                \label{eqn:nuhat-def}\hat{\nu}_{k}^{(n)}=\nu_{k}^{(n)}\mathcal{I}_{-\frac{k-1}{N}+n-1}.
            \end{equation}
            Using (\ref{eqn:magic-II}-\ref{eqn:magic-II-minus}), we can always define the coefficients $\Phi^{(n)}_{l}$, $\Psi^{(n)}_{l}$, $\tilde{\Phi}^{(n)}_{l}$, $\tilde{\Psi}^{(n)}_{l}$ 
             so that
            \begin{equation}\label{eqn:nu-hat-phi-psi}
                \hat{\nu}_{l+1+(n-1)N}^{(n)}=U^{-l}\left({\Phi}^{(n)}_{l}E_2^{(l)}+{\Psi}^{(n)}_{l}f_2^{(l)}(1-t)^{\frac{1}{2}}\right),
            \end{equation}
            and
            \begin{equation}\label{eqn:nu-hat-phi-psi-t}
                \hat{\nu}_{n N-l+1}^{(n)}=U^{l}\left(\widetilde{\Phi}^{(n)}_{l}E_2^{(l)}+\widetilde{\Psi}^{(n)}_{l}f_2^{(l)}(1-t)^{\frac{1}{2}}\right).
            \end{equation}
                Here the coefficients $\Phi^{(n)}_{l}$, $\Psi^{(n)}_{l}$, $\tilde{\Phi}^{(n)}_{l}$, $\tilde{\Psi}^{(n)}_{l}$ are rational functions of $q$, $\overline{\delta p_{N-1}}$ (see Appendix \ref{app:phipsi}), and 
            \begin{equation}
                U^N=(-1)^N(1-t)^{\frac{N}{2}}\left(-\frac{1}{f_2^{(1)}}\right)^{\frac{N}{2}}.
            \end{equation}
            \paragraph{Remark.} We note that if $N$ is even, $U^{-N}$ is a modular form  of $\Gamma_{(\frac{N}{N-2},\infty,\infty)}$. If $N$ is odd, we get a dependence on the branch of $\sqrt{f_2^{(1)}}$, so $U^{-N}$ may in principle change sign under some of the duality transformations. Understanding this phenomenon could be an interesting problem, but we shall not discuss it in the present paper. 
            \par 
            In the new notation, the equation determining $\overline{\delta p_{N-1}}$ (\ref{eqn:deltapL-bar}) takes the form
            
            \begin{equation}\label{eqn:deltapL-bar-new}
                \left(\widetilde{\Phi}^{(0)}_{1}E_2^{(1)}+\widetilde{\Psi}^{(0)}_{1}f_2^{(1)}(1-t)^{\frac{1}{2}}\right)=\mathcal{I}_{-\alpha-1}\mathcal{I}_{\alpha}=-f_2^{(1)}(1-t)^{-\frac{1}{2}}.
            \end{equation}
            Solving this equation perturbatively together with (\ref{eqn:tildephi},\ref{eqn:tildepsi}), we get  $\overline{\delta p_{N-1}}$ as a rational function of $E_2^{(1)}$, $\sqrt{f_2^{(1)}}$ and $q$. 
            \par Thus, every $\hat{\nu}_{k}^{(n)}$ can be expressed as a function of these variables.
            \par 
             Using these definitions, we rewrite (\ref{eqn:tau-l-nu}) as
            \begin{equation}\label{eqn:tau-l-nu-hat}
                2\pi \ii q\frac{\dd \tau^{(l)}_{\bm{m}}}{\dd q}= \frac{1+q}{1-q}\frac{\hat{\nu}^{(2)}_{N+2}}{N\hat{\nu}^{(1)}_{2}}\frac{f_2^{(l)}(1-t)^{-\frac{1}{2}}}{\hat{\nu}^{(1)}_{l+1}\hat{\nu}^{(1)}_{N-l+1}}.
            \end{equation}
            The right-hand side of this equation is a rational function of $f_2^{(l)}$, $E_2^{(l)}$, $E_2^{(1)}$, $\sqrt{f_2^{(1)}}$, and $q$. 
            \par 
            Let us consider the expansion of the right-hand side of (\ref{eqn:tau-l-nu-hat}) in powers of the mass scale \(\varepsilon\). Since it is a rational function homogeneous of degree \(-1\) in \(f_2^{(l)}\) and \(E_2^{(l)}\), while quasi-modular contributions always appear with positive powers of \(\varepsilon\) (this follows from straightforward algebra, which we omit), we may symbolically write
            \begin{equation}\label{eqn:EfEFmonomial}
                2\pi \ii q\frac{\dd \tau^{(l)}_{\bm{m}}}{\dd q}=\sum_{n_1,n_2\in \mathbb{N}_0, n_3\in \mathbb{Z}}\Omega_{n_1,n_2,n_3}(q,\varepsilon)\left(E_2^{(l)}\right)^{n_1}\left(f_2^{(l)}\right)^{1-n_1}\left(E_2^{(1)}\right)^{n_2}\left(f_2^{(1)}\right)^{\frac{n_3}2},
            \end{equation}
            where the coefficients $\Omega_{n_1,n_2,n_3}$ are rational functions of $q$, and power series in $\varepsilon$. Although the sum is formally infinite, only finitely many terms contribute in each order of $\varepsilon$. 
            \par 
           From (\ref{eqn:d-dtau}), we see that the operator $2\pi \ii q \frac{\dd }{\dd q}=\frac{\dd}{\dd \tau_{uv}}$ preserves the total degree of the modular and quasi-modular forms of each group. Thus, if we assume that
            \begin{equation}\label{eqn:DeltaTau}
                \Delta \tau^{(l)}= \tau_{\bm{m}}^{(l)}-\tau^{(l)}
            \end{equation}
            itself is a rational function of $f_2^{(l)}$, $E_2^{(l)}$, $E_2^{(1)}$, $\sqrt{f_2^{(1)}}$, and $q$, its expansion should admit the same form, \textit{i.e.} 
            \begin{equation}\label{eqn:EfEFmonomial-delta}
                \Delta \tau^{(l)}=\sum_{n_1,n_2\in \mathbb{N}_0, n_3\in \mathbb{Z}}\overline{\Omega}_{n_1,n_2,n_3}(q,\varepsilon)\left(E_2^{(l)}\right)^{n_1}\left(f_2^{(l)}\right)^{1-n_1}\left(E_2^{(1)}\right)^{n_2}\left(f_2^{(1)}\right)^{\frac{n_3}2},
            \end{equation}
            for some coefficients $\overline{\Omega}_{n_1,n_2,n_3}$. One can construct an algorithm to compute the coefficients $\overline{\Omega}$ from $\Omega$ at each order of $\varepsilon$ from (\ref{eqn:tau-l-nu-hat}) and (\ref{eqn:d-dtau}). We omit the details, because we shall eventually come up with a closed algebraic formula for $\Delta \tau^{(l)}$. 
            \par 
            Equation (\ref{eqn:EfEFmonomial-delta}) implies that
            \begin{equation} \label{eqn:assumption}
                \Delta \tau^{(l)}= \frac{1}{f_2^{(l)}} F^{(l)} \left({E_2^{(1)}},\sqrt{{f_2^{(1)}}},\frac{E_2^{(l)}}{f_2^{(l)}},q \right)
            \end{equation}
            for some rational function $F^{(l)}$.
            \paragraph{Remark.} We underline that the assumption that (\ref{eqn:tau-l-nu-hat}) can be integrated in terms of the modular objects $E_2^{(l)}$, $f_2^{(l)}$, $E_2^{(1)}$, $f_2^{(1)}$ and $q$ does not follow from our analysis. We have verified this explicitly up to order \(\varepsilon^{8N}\).

        \subsection{Anomaly equation and deformed coupling constants}
            \label{ssec:mod-anom}
            In this subsection, we derive a closed algebraic expression for the couplings $\tau_{\bm m}^{(l)}$. The resulting formula (\ref{eqn:best-for-tau-mass}) still depends on $\overline{\delta p_{N-1}}$ that have to be found from the nonlinear equation (\ref{eqn:deltapL-bar-new}), but does not require any further integration or ansatz.
            First, we assume that the ansatz (\ref{eqn:assumption}) is valid. Second, we assume that the modular anomaly equation of \cite{Pesandoetal2, Frauetal, Lerda} holds. We recall that this equation was itself derived from the $S$-duality of the theory together with the assumption that $\Delta \tau^{(l)}$ is a rational function of the modular objects, namely (\ref{eqn:assumption}).
            %deformed by the presence of masses more or less explicitly. However, integrating (\ref{eqn:tau-l-nu}) does not look possible. To proceed with the task, we recall the anomaly equation derived in \cite{Lerda}. 
            \par For simplicity, we restrict ourselves to the case $l= 2,\ldots,\lfloor\frac{N}{2}\rfloor$, but the final result is valid also for $l=1$. Then one of the two modular anomaly equations can be written as\footnote{It is easy to see that Eqn. (8.26) of \cite{Lerda} can be written in this form up to a numerical coefficient. The latter depends on the normalisation of $E_2^{(l)}$ which we have chosen differently with respect to \cite{Lerda}.} 
            \begin{equation}    \label{eqn:anomaly}
                \frac{\partial \Delta \tau^{(l)}}{\partial E_2^{(l)}}=-2\pi\ii(\Delta \tau^{(l)})^2 .
            \end{equation}
            % where
            % \begin{equation}
            %     \Delta \tau^{(l)}=\tau^{(l)}-(\Delta \tau^{(l)})_{\varepsilon =0}
            % \end{equation}
            % and $E_2^{(l)}$ a quasi-modular form of weight two and depth one of both the groups $\Gamma_{(\frac{2N}{N-2l},2,\infty)}$ (with Hauptmodul $t$) and $\Gamma_{(\frac{N}{N-2l},\infty,\infty)}$ (with Hauptmodul $q$) (see Appendix \ref{app:modularprop}). %Note that the coefficient in (\ref{eqn:anomaly}) depends on the exact choice of $E_2^{(l)}$, which is obviously not unique. We specify our notation in Appendix \ref{app:modularprop}; it is different from the one adopted in \cite{Lerda}.
            \par This equation can be immediately solved in the form
            \begin{equation}\label{eqn:mod-an-form-sol}
                \Delta \tau^{(l)}=\frac{1}{2\pi\ii}\frac{1}{E_2^{(l)}+C}.
            \end{equation}
            % where $C$ does not depend on $E_2^{(l)}$. We recall that (\ref{eqn:anomaly}) was derived under the assumption that $\Delta \tau^{(l)}$ is a rational function of $E_2^{(l)}$, $f_2^{(l)}$ and $q$. %In Appendix \ref{app:modularprop} we give an expression for the derivative with respect to $\tau_{\rm uv}$ acting on the functions of this kind (\ref{...}). Note that this operator, then applied to a monomial
            % \begin{equation*}
            %     \left(E_2^{(l)}\right)^{n_1}\left(f_2^{(l)}\right)^{n_2}\left(E_2^{(1)}\right)^{n_3}\left(f_2^{(1)}\right)^{n_4},
            % \end{equation*}
            % preserves the total degrees $n_1+n_2$ and $n_3+n_4$ of the modular and quasi-modular forms of each group. From (\ref{eqn:...}) we see that the formal power series expansion of $\frac{\dd \tau^{(l)}_{\bm{m}}}{\dd \tau_{\rm uv}}$ in terms of $\varepsilon$ consists exclusively of the monomials of such form with $n_1+n_2=1$, $n_3+n_4=0$. This means that the coupling constant is a function of the form
            \par Combining it with (\ref{eqn:assumption}), we get
            \begin{equation}\label{eqn:delta-tau-alg-1}
                \Delta \tau^{(l)}=\frac{1}{2\pi\ii}\frac{1}{E_2^{(l)}+f_2^{(l)}\Theta^{(l)}\left({E_2^{(1)}},\sqrt{f_2^{(1)}},q\right)},
            \end{equation}
            where $\Theta^{(l)}$ is a new rational function. This function could be found by taking derivative with respect to $\tau_{\rm uv}$ of the both sides of the equation above. This, however, still leads to a differential equation we are trying to avoid. 
            
           % \par  With the help of equation (\ref{eqn:anomaly}) we connect $\Delta \tau^{(l)}$ with the already found function  $\Delta  \dot{\tau}^{(l)}$ algebraically, avoiding any integration at all.
            \par From (\ref{eqn:anomaly}) and (\ref{eqn:assumption}) easily follows that  \begin{equation} \label{eqn:der1}
                \frac{\partial^2 \Delta \tau^{(l)}}{(\partial E_2^{(l)})^2} = -8\pi^2(\Delta \tau^{(l)})^3 ,
            \end{equation}
            \begin{equation} \label{eqn:der2}
                \frac{\dd}{\dd \tau_{\rm UV}} \frac{\partial \Delta \tau^{(l)}}{\partial E_2^{(l)}} =-4\pi\ii \Delta \tau^{(l)} \Delta  \dot{\tau}^{(l)},\ \frac{\dd}{\dd \tau_{\rm UV}} \frac{\partial^2 \Delta \tau^{(l)}}{(\partial E_2^{(l)})^2} =-24\pi^2 \left(\Delta \tau^{(l)}\right)^2 \Delta  \dot{\tau}^{(l)},
            \end{equation}
            \begin{equation} \label{eqn:der3}
                \frac{\partial \Delta \tau^{(l)}}{\partial f_2^{(l)}}= -\frac{\Delta \tau^{(l)}}{f_2^{(l)}}+2\pi\ii\frac{E_2^{(l)}}{f_2^{(l)}}  (\Delta \tau^{(l)})^2 
            \end{equation}
            and
            \begin{equation}
                     \frac{\partial^2 }{\partial (E_2^{(l)})^2} \frac{\dd \Delta \tau^{(l)}}{\dd \tau_{\rm uv}}-\frac{\dd }{\dd \tau_{\rm uv}} \frac{\partial^2 \Delta \tau^{(l)}}{\partial (E_2^{(l)})^2} 
                     =-8\pi \ii \frac{\partial^2\Delta \tau^{(l)}}{\partial E_2^{(l)}\partial f_2^{(l)}}-\frac{4\pi\ii}{f_2^{(l)}}\frac{\partial\Delta \tau^{(l)}}{\partial E_2^{(l)}}-8\pi\ii \frac{E_2^{(l)}}{f_2^{(l)}}\frac{\partial^2\Delta \tau^{(l)}}{(\partial E_2^{(l)})^2}.
            \end{equation}
            Combining all this together, we get
            \begin{equation}
                \frac{\partial^2 }{\partial (E_2^{(l)})^2} \frac{\dd \Delta \tau^{(l)}}{\dd \tau_{\rm uv}}+24\pi^2 \left(\Delta \tau^{(l)}\right)^2 \Delta  \dot{\tau}^{(l)}=\frac{24\pi^2}{f_2^{(l)}}\left(\Delta \tau^{(l)}\right)^2.
            \end{equation}
            Taking (\ref{eqn:ourf2}) into account, we can write it as
            \begin{equation} \label{eqn:tau-from-anomaly}
                \left(\Delta \tau^{(l)}\right)^2=\frac{1}{(2\pi\ii)^2}\frac{1}{6 \dot{\tau}^{(l)}_{\bm m}}\frac{\partial^2 \dot{\tau}^{(l)}_{\bm m}}{(\partial E_2^{(l)})^2},
            \end{equation}
                where the dot stands for $\frac{\dd}{\dd \tau_{\rm uv}}$.
                This is the desired equation, which allows us to extract all orders of $\Delta \tau^{(l)}$ from (\ref{eqn:tau-l-nu-hat}) without integration. 
                \par 
                Note that the only factors in (\ref{eqn:tau-l-nu-hat}) which depend on $E_2^{(l)}$ are the functions $\hat{\nu}_{l+1}^{(1)}$ and $\hat{\nu}_{N-l+1}^{(1)}$. So, (\ref{eqn:tau-from-anomaly}) can be rewritten as
                \begin{equation}
                    \left(\Delta \tau^{(l)}\right)^2=\frac{1}{(2\pi\ii)^2}\frac{(E_2^{(l)})^2+\left(\Xi_l+\widetilde{\Xi}_l\right)E_2^{(l)}f_2^{(l)}+\frac{1}{3}\left(\Xi_l^2+\Xi_l\widetilde{\Xi}_l+\widetilde{\Xi}_l^2\right)(f_2^{(l)})^2}{(E_2^{(l)}+f_2^{(l)}\Xi_l)^2(E_2^{(l)}+f_2^{(l)}\widetilde{\Xi}_l)^2},
                \end{equation}
            where 
            \begin{equation}
                 \Xi_l=(1-t)^{\frac{1}{2}}\frac{\Psi^{(1)}_l}{\Phi^{(1)}_l},\ \widetilde{\Xi}_l=(1-t)^{\frac{1}{2}}\frac{\widetilde{\Psi}^{(1)}_l}{\widetilde{\Phi}^{(1)}_l},
            \end{equation}
            and we used the notation introduced in (\ref{eqn:nu-hat-phi-psi}-\ref{eqn:nu-hat-phi-psi-t}). This form clearly contradicts (\ref{eqn:delta-tau-alg-1}) unless $\Xi_l=\widetilde{\Xi}_l=\Theta_l$. Unfortunately, at this moment we are not able to prove that this holds in general, but we have verified this equality up to $\varepsilon^{8N}$. Assuming $\Xi_l=\tilde{\Xi}_l$, we can use (\ref{eqn:delta-tau-alg-1}) to get the right sign of the square root, finally arriving at
            \begin{equation}\label{eqn:best-for-tau-mass}
                \Delta \tau^{(l)}=\frac{1}{2\pi \ii}\frac{\Phi^{(1)}_l}{\Phi_l^{(1)} E_2^{(l)}+\Psi^{(1)}_l f_2^{(l)}}.
            \end{equation}
            and the coefficients $\Phi_l^{(1)}$ and $\Psi_l^{(1)}$ are given by (\ref{eqn:Phi}) and (\ref{eqn:Psi}).
            %  The coefficient $\Theta_l$ can be computed using a computer algebra system and the formula
            % \begin{equation},,, y} H_{n-1,l\alpha-1}U^{-n N}s^{Nn-1}G_1(s)V(s)^{-\frac{l}{N}-n}}{{\rm Res_{s=0}}\sum_{n} H_{ n,l\alpha}U^{-n N}s^{N n-1}G_1(s)V(s)^{-\frac{l}{N}-n}}
            % \end{equation}       
            % \begin{eqnarray}
            %     \Phi_l^{(1)}=-\frac{1}{N}\sum_{n} {\rm Res_{s=0}} H_{n-1,l\alpha-1}U^{-n N}s^{Nn-1}G_1(s)V(s)^{-\frac{l}{N}-n}=\\
            %     -\frac{T}{NA^NU^N}+\frac{2NT'-N T^2 -2 T(-A)^N (N+l)\overline{\delta p_{N-1}}}{2N^2A^{2N}U^{2N}}+\mathcal{O}(\varepsilon^{3N}),\nonumber
            % \end{eqnarray}
            % \begin{eqnarray}
            %     \Psi_l^{(1)}=-\frac{1}{N}\sum_{n} {\rm Res_{s=0}} H_{ n,l\alpha}U^{-n N}s^{N n-1}G_1(s)V(s)^{-\frac{l}{N}-n}=\\
            %     -\frac{2}{l(1-t)}-\frac{T+2(-A)^N \overline{\delta p_{N-1}}}{N A^NU^N(1-t) }+\mathcal{O}(\varepsilon^{2N})\nonumber.
            % \end{eqnarray}
            \par In the limit of zero masses, $\Phi_l^{(1)}$ tends to zero, so the first corrections do not contain the quasi-modular forms $E_2^{(l)}$ in agreement with \cite{Lerda}:
            \begin{equation}
                \Delta \tau^{(l)}=\frac{\sqrt{1-t}}{2\pi \ii}\frac{l T}{2N A^N U^N f_2^{(l)}}+\mathcal{O}(\varepsilon^{2N}).
            \end{equation}

        \par As a final remark, we note that (\ref{eqn:mod-an-form-sol}) closely resembles (\ref{eqn:E2-higher}). This suggests that, in the absence of any dependence of the coefficient $\Theta^{(l)}$ on ${E_2^{(1)}}$ and $\sqrt{f_2^{(1)}}$, the massive coupling constants $\tau^{(l)}_{\bm m}$ would be equimodular functions of $\tau^{(l)}$ (see Appendix \ref{app:modularprop}); \textit{i.e.}, they would obey exactly the same transformation law as $\tau^{(l)}$. However, the appearance of ${E_2^{(1)}}$ and $\sqrt{f_2^{(1)}}$ makes the transformation properties more complicated, since they involve not only $\tau^{(l)}$, but also $\tau^{(1)}$ through (\ref{eqn:f2tranform},\ref{eqn:E2def}). Thus, $\tau^{(1)}$ continues to play a distinguished role in the massive special vacuum.
           
    \section{Summary}
    In this paper, we have found a universal form of the effective coupling matrix in the special vacuum
    \begin{equation}\label{eqn.gen-tau-sum}
          \tau_{uv}^{\mathrm{IR}}=
            \frac{1}{ N}\sum_{l=1}^{N-1}\tau^{(l)}    \left(\cos\left(\frac{2\pi (u-v)l}{N}\right)-\cos\left(\frac{2\pi u l}{N}\right)-\cos\left(\frac{2\pi v l}{N}\right)+1\right),
    \end{equation}
    where $\tau^{(l)}=\tau^{(N-l)}$ ($l=1,\ldots,N-1$) are the $\lfloor\frac{N}{2}\rfloor$ couplings. 
    \par 
    The universal form of the coupling matrix follows solely from the residual Weyl symmetry (\textit{i.e.} the unbroken subgroup of the Weyl group) and dimensional principles. 
    \par 
    In the massless case, we have found a simple explicit formula for all the couplings,
           \begin{eqnarray} 
            2\pi \ii \tau^{(l)}=-\frac{\Gamma\left(\frac{l}{N}\right)^2}{\Gamma\left(\frac{2l}{N}\right)}\frac{{{}_2F_1\left(\frac{l}{N},\frac{l}{N},\frac{2l}{N},1-q\right)}}{{}_2F_1\left(\frac{l}{N},\frac{l}{N},1,q\right)}+\pi\left(\cot\left(\frac{\pi l}{N}\right) +\ii\right)=\\
            -\frac{\Gamma\left(\frac{l}{N}\right)^2}{\Gamma\left(\frac{2l}{N}\right)}2^{\frac{l}{N}} \frac{{}_2F_1\left(\frac{l}{2N},\frac{N+l}{2N},\frac{N+2l}{2N};1-t\right)}{{}_2F_1\left(\frac{l}{2N},\frac{N+l}{2N},1;t\right)} + \pi\left(\cot\left(\frac{\pi l}{N}\right) +\ii\right),
        \end{eqnarray}
        where $t=\frac{4}{q+\frac{1}{q}+2}$ s invariant under $S$-duality. For the massless theory, we also find a compact, manifestly symmetric expression for the prepotential,
      \begin{eqnarray}\label{eqn:genF-sum}
            \mathcal{F}=2\pi \ii \left(\tau_{\rm IR} w_1 -\frac{1}{2N}\sum_{l=2}^{N-2}\frac{w_{l} w_{N-l}}{w_{N-1}}(\tau_{\rm IR}-\tau^{(l)})\right)=\frac{2 \pi \ii}{2N}\sum_{l=1}^{N-1}\frac{w_{l} w_{N-l}}{w_{N-1}}\hat{\tau}^{(l)},
        \end{eqnarray}
        where 
        \begin{equation}
            \hat{\tau}^{(l)}=\tau^{(l)}-\tau_{\rm IR}(1-2N \delta_{l,1}).
        \end{equation}
        The dual variables are given by
        \begin{equation}
            a_u^D=-\frac{2\pi\ii}{\sqrt{N}}\tau^{(1)}a_u
        \end{equation}
        \par
   The couplings $\tau^{(l)}$ are subject to the $S$-duality group action according to
   \begin{equation}
       S:\ \tau^{(l)}\to -\frac{1}{4\sin\left(\frac{\pi l}{N}\right)^2  \tau^{(l)}},\quad T:\ \tau^{(l)}\to \tau^{(l)}+1. 
   \end{equation}
   Up to a shift and rescaling, these two transformations generate the (generalised) triangle group $\Gamma_{(\frac{2N}{N-2l},2,\infty)}$. Accordingly, the inverse function $\tau^{(l)}\mapsto t$ can be interpreted as a Hauptmodule of this group. The original coupling $q$ is also the Hauptmodule of each of the couplings, but with respect to the $q$-preserving subgroup $\Gamma_{(\frac{N}{N-2l},\infty,\infty)}\subset \Gamma_{(\frac{2N}{N-2l},2,\infty)}$ generated by the transformations $T$ and $STS$. 
   \par 
   We also provide an algorithm of computing the corrections to the couplings in the massive special vacuum; see Subsection \ref{ssec:diffeq} and Appendix \ref{sub-app:nukn}. Moreover, assuming that the massive corrections $\Delta \tau^{(l)}=\tau^{(l)}_{\bm{m}}-\tau^{(l)}$ in any order of mass are rational functions of the modular and quasimodular forms, we managed to derive their closed form
    \begin{equation}
                \Delta \tau^{(l)}=\frac{1}{2\pi \ii}\frac{\Phi^{(1)}_l}{\Phi_l^{(1)} E_2^{(l)}+\Psi^{(1)}_l f_2^{(l)}}.
            \end{equation}
   Here $f_2^{(l)}$, $E_2^{(l)}$ are respectively the modular and quasi-modular forms of the $\Gamma_{(\frac{N}{N-2l},\infty,\infty)}$ group, defined in Appendix \ref{app:modularprop}, and $\Phi$, $\Psi$ are rational functions of $f_2^{(1)}$, $E_2^{(1)}$, $q$, $A$, and the masses, computed in Appendix \ref{app:phipsi}. We have verified that this form and the underlying conjecture are valid up to the eighth non-trivial order of mass corrections.
   \par 
   The expression for the dual variables in this case is less direct. Namely, we have
   \begin{equation}
       a_u^D=-\frac{2\pi\ii}{\sqrt{N}}\tau_{\mathrm{IR},\bm{m}}a_u,
   \end{equation}
   where
   \begin{equation}
       \frac{\partial}{\partial A}A \tau_{\mathrm{IR},\bm{m}} = \tau_{\bm{m}}^{(1)}.
   \end{equation}
   \par 
   Although the results above treat the $\lfloor N/2\rfloor$ couplings, and their associated triangle groups, in a seemingly symmetric manner, $\tau^{(1)}$ in fact plays a distinguished role. This asymmetry is reflected in three important properties:
   \begin{enumerate}
       \item Only $\tau^{(1)}$ is relevant for the computation of the dual variables $a_u^D$ in the special vacuum.
       \item The massive corrections $\Delta\tau^{(l)}$ for ever $l$ depend on $f_2^{(1)}$, $E_2^{(1)}$, the (quasi-) modular forms of $\tau^{(1)}$.
       \item Last but not least, only $\tau^{(1)}$ survives in the limit $A\to \infty$ (or equivalently $w_{N-1}\to \infty$). 
   \end{enumerate}
   As a consequence of the latter, only $\tau^{(1)}$ enters in the Zamolodchikov-like relation, recurrently computing the whole equivariant partition function for generic vacuum expectation values $a_u$ and any masses, as a formal power series in term of only one exponentiated coupling $q_{\mathrm{IR}}=e^{2\pi\ii \tau^{(1)}|_{A\to \infty}}$. Thus, even though the special vacuum description involves $\lfloor N/2\rfloor$ distinct effective couplings, only one of them is required as the fundamental expansion parameter of the full partition function. The remaining $\lfloor N/2\rfloor-1$ couplings are therefore not independent from the viewpoint of this expansion; rather, their information is encoded in the coefficients of the formal power series in $q_{\mathrm{IR}}$.
   The reason behind of this unfairness is, of course, the non-zero value of $w_{N-1}$.
   \par Taken together, these results reveal a surprisingly rigid structure underlying the effective couplings in the special vacuum: their universal form is fixed by symmetry, their massless dependence is governed by triangle-group modularity, and their massive corrections are controlled by the modular anomaly equation, leaving only two coefficients for each coupling, depending on just two modular objects. At the same time, the distinguished role of $\tau^{(1)}$ shows that this apparently multi-coupling description ultimately possesses a single preferred modular parameter.

    \section*{Acknowledgements}
    
        E.S. would like to thank Marialuisa Frau for discussions during the project.
        \par Research of E.S. is partially supported by the MUR PRIN contract 2020KR4KN2 "String Theory as a bridge between Gauge Theories and Quantum Gravity" and by the INFN project ST\&FI "String Theory \& Fundamental Interactions".

    \appendix

    \section{Modular and equimodular functions, modular and quasi modular forms} \label{app:modularprop}
        
        In this appendix, we define the functions $f_2^{(l)}$ and $E_2^{(l)}$ and study their properties.
        \par The main motivation for their introduction is the fact that, whenever $\frac{N}{N-2l}\in\mathbb{N}$, these two functions are a weight two modular form and weight two quasi-modular form (see below). Moreover, any modular or quasi-modular form is a rational function of  $f_2^{(l)}$, $E_2^{(l)}$ and $q$. This fact allows to derive various properties of $f_2^{(l)}$ and $E_2^{(l)}$. 
        For general values of $N$ and $l$, when the modular invariance is more subtle, these properties still hold. For this reason, it is very convenient to express the corrections to coupling constants through $f_2^{(l)}$ and $E_2^{(l)}$. Last but not least, the modular anomaly equation of \cite{Lerda} is written in terms of the formal derivative $\frac{\partial }{\partial E_2^{(l)}}$.

        \subsection{Basic formulas}
            We begin by recalling the basic properties of the Hypergeometric function, applied to the special case
            \begin{equation}
                I_\alpha = {}_2 F_1 \left(-\frac{\alpha}{2},\frac{1-\alpha}{2},1;t\right).
            \end{equation}
            We shall use the following relations (see, \textit{e.g.}, \cite{NIST}):
            \begin{equation} \label{eqn:prop1}
                I_{-\alpha}=(1-t)^{\frac{1}{2}-\alpha}I_{\alpha-1},
            \end{equation}
            \begin{equation} \label{eqn:prop2}
                I_{\alpha-1}=I_{\alpha}-\frac{2t}{\alpha}\dot{I}_{\alpha},
            \end{equation}
            \begin{equation} \label{eqn:prop3}
                \left( \frac{\alpha+1}{2}(1-t)+\frac{2\alpha+1}{2}t\right)I_\alpha+t(1-t)\dot{I}_\alpha=\frac{\alpha+1}{2}I_{\alpha+1}.
            \end{equation}

        \subsection{Modular form}
            Following \cite{BykovSysoeva}, we introduce
            \begin{equation} \label{eqn:ourf2}
                f_2^{(l)} =- \frac{\dd \tau_{\rm uv}}{\dd \tau^{(l)}}=-(1-t)^{-\alpha l} I_{l \alpha} ^2=-(1-t)^{1+\alpha l} I_{-l \alpha-1} ^2=-I_{-l \alpha-1}I_{ l\alpha}(1-t)^{\frac{1}{2}}. %{}_2F_1\left( -\frac{\alpha l}{2}, \frac{1-\alpha l}{2}, 1; t\right)^2 .
            \end{equation}
            Whenever $N-2l$ divides $N$, $f_2^{(l)}$  is a weight two modular form for the group $\Gamma_{(\frac{N}{N-2l},\infty,\infty)}$ \cite{Doran}, \textit{i.e.}, as a function of $\tau^{(l)}$, it transforms as
            \begin{equation}
                f_2^{(l)} \left( \frac{a \tau^{(l)} +b}{c \tau^{(l)} +d} \right) = (c \tau^{(l)} + d)^2 f_2(\tau^{(l)}),
            \end{equation}
            where 
            \begin{equation}
                \left(\begin{matrix}
                    a & b \\
                    c & d
                \end{matrix}\right) \in \Gamma_{(\frac{N}{N-2l},\infty,\infty)}.
            \end{equation}
            We note that for the larger group $\Gamma_{(\frac{2N}{N-2l},2,\infty)}$, according to \cite{Doran}, the generating modular form is
            \begin{equation} \label{eqn:f2tranform}
                 \tilde{f}_2^{(l)}= f_2^{(l)}t(1-t)^{\frac{1}{2}}.
            \end{equation}
            \textbf{Remark.}
            While the modular function (for both groups) $t$ does not affect the modular properties, the factor $(1-t)^{\frac{1}{2}}=\frac{1-q}{1+q}$ changes its sign under the $S$-transformation. Thus,  $f_2^{(l)}$ is not a modular form for $\Gamma_{(\frac{2N}{N-2l},2,\infty)}$, and instead transforms as 
            \begin{equation}
                f_2^{(l)} \left( \frac{a \tau^{(l)} +b}{c \tau^{(l)} +d} \right) = \pm (c \tau^{(l)} + d)^2 f_2(\tau^{(l)}).
            \end{equation}
            We also recall that in the cases $N/l=2,3,4$, both groups $\Gamma_{(\frac{2N}{N-2l},2,\infty)}$ and $\Gamma_{(\frac{N}{N-2l},\infty,\infty)}$ are arithmetic. The terminology commonly used for the arithmetic groups is different from the one used for the triangle groups. In particular, in the arithmetic modular groups terminology, the function $f_2^{(l)}$ is a modular form, but with a nontrivial character of $\frac{\Gamma_{(\frac{2N}{N-2l},2,\infty)}}{\Gamma_{(\frac{N}{N-2l},\infty,\infty)}}\cong \mathbb{Z}_2$.

        \subsection{Equimodular functions and quasi-modular forms}  
            From (\ref{eqn:vl-G}), one can see that functions $I_{ \alpha+n}$ for arbitrary large integer $n$ can appear in the higher order corrections. In this light, it is natural to formally consider $\tau^{(l+nN)}$ with $n\in \mathbb{Z}$.  From (\ref{eqn:S-transform}) and (\ref{eqn:T-transform}) we see that $\tau^{(l+nN)}$ transform in exactly the same way as does $\tau^{(l)}$. It is natural to say that $\tau^{(l+nN)}$ is an equimodular function of $\tau^{(l)}$. This fact has an interesting consequence. Let us consider $E_2^{(l)}$ defined by 
            \begin{equation}
                \tau^{(l+N)}-\tau^{(l)}=\frac{1}{2 \pi \ii }\frac{1}{E_2^{(l)}}.
            \end{equation}
            Its transformation law is 
            \begin{equation}\label{eqn:E-trans-der}
            \frac{1}{2 \pi \ii }\frac{1}{E_2^{(l)}}\to    \frac{a \tau^{(l+N)} +b}{c \tau^{(l+N)} +d}-\frac{a \tau^{(l)} +b}{c \tau^{(l)} +d}=\frac{1}{2 \pi \ii }\frac{1}{(c \tau^{(l)} +d)^2 E_2^{(l)}+\frac{c}{2 \pi \ii}(c \tau^{(l)} +d)},
            \end{equation}
            or
            \begin{equation}  \label{eqn:E2def}
                E_2 \left( \frac{a \tau^{(l)} +b}{c \tau^{(l)} +d} \right) = (c \tau^{(l)} + d)^2 E_2(\tau^{(l)}) + \frac{c}{2\pi\ii} \cdot (c \tau^{(l)} + d).
            \end{equation}
            We recognise that $E_2^{(l)}$ is a depth one, weight two quasi-modular form \cite{Doran,Zagier} for both $\Gamma_{(\frac{N}{N-2l},\infty,\infty)}$ and $\Gamma_{(\frac{2N}{N-2l},2,\infty)}$. It can be presented in a simple form
            
            \begin{equation}  \label{eqn:ourE2}
                E_2^{(l)}= - \frac{l \alpha}{2} (1-t)^{-l \alpha + \frac{1}{2} }I_{l \alpha} I_{l \alpha-1} .
            \end{equation}
             We only highlight the main steps of the proof of this fact. The crucial one is to write (\ref{eqn:ourf2}) as 
            \begin{equation}
                \frac{1}{ t(1-t)^{\frac{1}{2}}}\left(\frac{\dd}{\dd t}\left(
               \frac{\Gamma\left(-l\alpha\right)^2}{\Gamma\left(-2l\alpha\right)}2^{-l\alpha} \frac{{}_2F_1\left(\frac{1}{2N},\frac{1+l\alpha}{2},\frac{1}{2}+l\alpha;1-t\right)}{{}_2F_1\left(\frac{l\alpha}{2},\frac{1+l\alpha}{2},1;t\right)}\right) \right)^{-1}=(1-t)^{-\alpha l} I_{l \alpha} ^2.
            \end{equation}
            Then one has to compute the derivative in the left-hand side, and get rid of the hypergeometric functions with the argument $1-t$ by expressing them via $\tau^{(l)}$ and $\tau^{(l+N)}$. The rest is a tedious  technical computation based on standard properties of the hypergeometric functions which we omit.
            \par 
            The field of rational functions of $q$, $E_2^{(l)}$ and $f_2^{(l)}$ is closed under the derivative with respect to $\tau_{\rm{uv}}$:
            \begin{equation}
            \frac{\dd f_2^{(l)}}{\dd \tau_{\rm uv}}=-2\pi \ii \left( l \alpha(1-t)^{-\frac{1}{2}} f_2^{(l)}+2E_2^
            {(l)}\right)
            \end{equation}
            \begin{equation}
            \frac{\dd E_2^{(l)}}{\dd \tau_{\rm uv}}=2\pi\ii\left(\frac{(l\alpha)^2}{4}f_2^{(l)}-\frac{\left(E_2^{(l)}\right)^2}{f_2^{(l)}}\right)
            \end{equation}
            \begin{equation}
                \frac{\dd t}{\dd \tau_{\rm uv}} = 2\pi \ii q\frac{\dd t}{\dd q}=-2\pi\ii \, t (1-t)^{\frac{1}{2}}.
            \end{equation}
            This allows us to write 
            \begin{equation}\label{eqn:d-dtau}
                \frac{\dd }{\dd \tau_{\rm uv}}=\sum_{l=1}^{\lfloor \frac{N}{2}\lfloor}\frac{\dd f_2^{(l)}}{\dd \tau_{\rm uv}} \frac{\partial}{\partial f_2^{(l)}}+\sum_{l=1}^{\lfloor \frac{N}{2}\rfloor}\frac{\dd E_2^{(l)}}{\dd \tau_{\rm uv}} \frac{\partial}{\partial E_2^{(l)}}+\frac{\dd t}{\dd \tau_{\rm uv}} \frac{\partial}{\partial t}.
            \end{equation}
            Here we assume that the derivative acts on a rational function of $t$, $f_2^{(l)}$ and $E_2^{(l)}$, and $ \frac{\partial}{\partial f_2^{(l)}}$ and $\frac{\partial}{\partial E_2^{(l)}}$ are formal partial derivatives with respect to the corresponding arguments\footnote{In reality, we allow also dependence on $\sqrt{f_2^{(1)}}$. This does not change the form of the operator}.

        \subsection{Further shifts of \texorpdfstring{$l$}{l}}
            In the previous subsections we concentrated on $\tau^{(l+N)}$. Further shifted coupling constants $\tau^{(l+nN)}$ are equimodular functions as well, so 
            \begin{equation}
                \frac{1}{2\pi\ii}\frac{1}{\tau^{(l+n N)}-\tau^{(l)}}
            \end{equation}
            must have the same transformation properties as does $E_2^{(l)}$ (at least for integer values of $\frac{N}{N-2l}$). Given that in the case of the triangular groups the quasi-modular form $E_2^{(l)}$ is uniquely determined by its transformation law up to a modular piece \cite{Doran}, there should be a relation
             \begin{equation}\label{eqn:E2-higher}
               \frac{1}{2\pi\ii}\frac{1}{\tau^{(l+n N)}-\tau^{(l)}}  =E_2^{(l)}+f_2^{(l)}(1-t)^{\frac{1}{2}}R_n^{(l)},
             \end{equation}
             where $R_n^{(l)}$ is a rational function of $t$. 
             \par 
             Before explaining how (\ref{eqn:E2-higher}) can be proved without relying on the modularity, let us write an alternative expression for the left-hand sides. 
            It is easy to see that, for positive integer $n$,
            \begin{equation}\label{eqn:taun-taul}
                \tau^{(l+nN)}-\tau^{(l)}=\frac{1}{2\pi\ii}\left(\frac{1}{E_2^{(l)}}+\cdots+\frac{1}{E_2^{(l+(n-1)N)}}\right)
            \end{equation}
            and 
            \begin{equation}\label{eqn:taul-taun}
                \tau^{(l)}-\tau^{(l-nN)}=\frac{1}{2\pi\ii}\left(\frac{1}{E_2^{(l-1)}}+\cdots+\frac{1}{E_2^{(l-nN)}}\right).
            \end{equation}
            Setting $n=2$ and combining this with (\ref{eqn:E2-higher}), we get
            \begin{equation}\label{eqn:E2-shift1}
                \frac{1}{2\pi\ii}\left(\frac{1}{E_2^{(l)}}+\frac{1}{E_2^{(l+N)}}\right)=\frac{1}{2\pi\ii}\frac{1}{E_2^{(l)}+f_2^{(l)}R_1^{(l)}}.
            \end{equation}
            This relation can be proven using (\ref{eqn:prop1}-\ref{eqn:prop3}). 
            Then we can proceed by induction for any positive $n$. Indeed, assuming that (\ref{eqn:E2-higher}) holds for $n=n_0-1$ (and arbitrary $l$), we write
            \begin{equation}
                \tau^{(l)}-\tau^{(l+n_0 N)}=\tau^{(l)}-\tau^{(l+1)}+\tau^{(l+1)}-\tau^{(l+n_0 N)}=\frac{1}{2\pi\ii}\frac{1}{E_2^{(l)}}+\frac{1}{2\pi\ii}\frac{1}{E_2^{(l+N)}+f_2^{(l+N)}(1-t)^{\frac{1}{2}}R_{n_0-1}^{(l+N)}}.
            \end{equation}
            Substituting $E_2^{(l+N)}$ from (\ref{eqn:E2-shift1}) and gathering two fractions to one, one gets (\ref{eqn:E2-higher}) for $n=n_0$. The case of negative $n$ can be treated similarly. Note that from the argument above one can in principle extract a recurrence formula for $R_n^{(l)}$. We do not present it here, because a better, direct formula (\ref{eqn:Rnl}) will be given in the next subsection. The existence of such a rational function $R_n^{(l)}$ is, however, important for derivation of that direct formula.

        \subsection{Universal formula for shifted functions}  \label{app:universalf}
            In this subsection, we present and prove a convenient explicit formula, computing $I_{l\alpha\pm n}$ for any $n\in\mathbb{Z}$ as a linear combination of $E_2^{(l)}$ and $f_2^{(l)}$, with rational functions of $q$ as coefficients. Namely, we claim that
            \begin{equation}\label{eqn:magic-II}
                \mathcal{I}_{l\alpha}\mathcal{I}_{l\alpha-n}(1-t)^{n-l\alpha-\frac{1}{2}}=H_{n,l\alpha}E_2^{(l)}+\frac{\alpha}{2}H_{n-1,l\alpha-1}f_2^{(l)}(1-t)^{\frac{1}{2}},
            \end{equation}
            \begin{equation}\label{eqn:magic-II-minus}
                \mathcal{I}_{l\alpha}\mathcal{I}_{-l\alpha-n-1}=H_{-n,l\alpha}E_2^{(l)}+\frac{\alpha}{2}H_{-n-1,l\alpha-1}f_2^{(l)}(1-t)^{\frac{1}{2}},
            \end{equation}
            where
            \begin{equation}\label{eqn:Hnalpha}
                H_{n,\alpha}=2 \, {\rm Sign}(n)\sum_{k=\frac{n-|n|}{2}}^{\frac{n+|n|}{2}-1}\frac{\mathcal{I}_{k}\mathcal{I}_{n-k-1}}{\alpha-k}
            \end{equation}
            with ${\rm Sign}(0)=0$.
            
            % \begin{equation}
            %     \label{eqn:Hnalpha}H_{n,\alpha}=2\sum_{k=0}^{n-1}\frac{\mathcal{I}_{k}\mathcal{I}_{n-k-1}}{\alpha-k}, \ n>0
            % \end{equation}
            % \begin{equation}
            %     H_{n,\alpha}=-2\sum_{k=n}^{-1}\frac{\mathcal{I}_{k}\mathcal{I}_{n-k-1}}{\alpha-k}, \ n<0
            % \end{equation}
            % \begin{equation}
            %     H_{0,\alpha}=0.
            % \end{equation}
            Here for positive integer $k$ the functions $\mathcal{I}_k$ turn out to be polynomials of $t$,
            \begin{equation}
                \mathcal{I}_{k}=\sum_{j=0}^{\lfloor\frac{k}{2}\rfloor}\frac{k!}{(j!)^2(k-2j)!}\left(\frac{t}{4}\right)^j
            \end{equation}For $k$ being a negative integer, we use (\ref{eqn:prop1}) with $\alpha=k$
            % \begin{equation}\label{eqn:I-k}
            %     \mathcal{I}_{-k}=(1-t)^{k-\frac{1}{2}}\mathcal{I}_{k-1},
            % \end{equation}
            giving
            \begin{equation}
                H_{-n,\alpha}=H_{n,-\alpha-1}(1-t)^{-n}.
            \end{equation}
            We also have
            \begin{equation}
                H_{n,\alpha}=-H_{n,n-\alpha-1},\ 
                H_{-n,-\alpha}=H_{n,\alpha+n}(1-t)^{-n}
            \end{equation}
            We conclude that for any integer value of $n$, $H$ is a rational function of $t$.
            \par 
            One can note that (\ref{eqn:magic-II}) essentially presents $\mathcal{I}_{l\alpha-n}$ as a linear combination of $\mathcal{I}_{l\alpha}$ and $\mathcal{I}_{l\alpha-1}$ with polynomials of $t$ as coefficients. For any finite value of $n$, expressions of this type can be straightforwardly derived from the standard relations. However, this general form seems much more difficult to achieve. Here we present some hints how it can be done using the modular properties of $E_2^{(l)}$ and $f_2^{(l)}$ as a guidance. It is enough to assume to prove (\ref{eqn:magic-II}) for $n>0$. The case $n=0$ is straightforward, and the rest follows by (\ref{eqn:prop1}).
            \par 
            Our starting point will be (\ref{eqn:E2-higher}). We define $H_{n,l\alpha}$ as
            \begin{equation}\label{eqn:H2b}
              H_{n,l\alpha}:=   \mathcal{I}_{l\alpha}\mathcal{I}_{l\alpha-n}(1-t)^{n-l\alpha-\frac{1}{2}}\left(  \tau^{(l)}-\tau^{(l+nN)}\right).
            \end{equation}
            We note that both $\tau^{(l)}$ and $\tau^{(l+nN)}$ depend on $l$ and $N$ exclusively via $l\alpha=-\frac{l}{N}$. This remark will be crucial when we formally consider $l\alpha$ as a complex parameter.
            \par 
            At this point it is not clear neither why $H_{n,l\alpha}$ is a rational function of $t$, nor why (\ref{eqn:magic-II}) holds. We start by arguing the former. We observe that $H_{n,l\alpha}$ defined by (\ref{eqn:H2b}) 
            is a modular function, i.e. a rational function of $t$. First of all, note that for even $n$ we can write
            \begin{eqnarray}
                 \label{eqn:II-even}\mathcal{I}_{l\alpha}\mathcal{I}_{l\alpha-n}(1-t)^{n-l\alpha-\frac{1}{2}}=\\(\ldots)\frac{E_2^{(l+(n-1)N)}}{E_2^{(l+(n-2)N)}}\frac{E_2^{(l+(n-3)N)}}{E_2^{(l+(n-4)N)}}\cdots \frac{E_2^{(l+N)}}{E_2^{(l)}}f_2^{(l)},
            \end{eqnarray}
            while for odd $n$
            \begin{eqnarray}
                 \label{eqn:II-odd}\mathcal{I}_{l\alpha}\mathcal{I}_{l\alpha-n}(1-t)^{n-l\alpha-\frac{1}{2}}=\\(\ldots)\frac{E_2^{(l+(n-1)N)}}{E_2^{(l+(n-2)N)}}\frac{E_2^{(l+(n-3)N)}}{E_2^{(l+(n-4)N)}}\cdots \frac{E_2^{(l+2N)}}{E_2^{(l+N)}}E_2^{(l)}.
            \end{eqnarray}
            Here $(\ldots)$ stands for a rational function of $t$.
            In both cases, this, together with (\ref{eqn:E2-higher}) and(\ref{eqn:taun-taul}-\ref{eqn:taul-taun}), implies that $H_{n,l \alpha}$ defined by (\ref{eqn:H2b}) is a rational function of $t$, $f_2^{(\ell)}$ and $E_2^{(\ell)}$. Moreover, if we introduce a grading, by assigning the degree 0 to $t$ and the degree $1$ to $f_2^{(\ell)}$ and $E_2^{(\ell)}$, then $H_n^{(\ell)}$ is a  homogenous rational function of degree zero. This implies that we need only to show that it does not depend on $E_2^{(\ell)}$. For that we compute the formal derivative $\frac{\partial}{\partial E_2^{(l)}}$.  From (\ref{eqn:E2-higher}), we get
            \begin{equation}\label{eqn:tau-over-E2}
               \frac{\partial}{\partial E_2^{(l)}}\left(  \tau^{(l)}-\tau^{(l+n N)}\right)=-2\pi \ii \left(  \tau^{(l)}-\tau^{(l+n N)}\right)^2.
            \end{equation}
            From this and (\ref{eqn:taul-taun}) follows
            \begin{equation}\label{eqn:E2-shift-der}
                \frac{1}{E_2^{(l+nN)}}\frac{\partial E_2^{(l+nN)}}{\partial E_2^{(l)}}=2\pi\ii \left( \tau^{(l)}-\tau^{(l+n N)}\right)+2\pi\ii \left(  \tau^{(l)}-\tau^{(l+(n-1)N)}\right).
            \end{equation}
            Acting by $\frac{\partial}{\partial E_2^{(l)}}$ on both hand sided (\ref{eqn:II-even}-\ref{eqn:II-odd}) and substituting (\ref{eqn:E2-shift-der}), we get
            \begin{equation}
                \frac{\partial }{\partial E_2^{(l)}}   \mathcal{I}_{l\alpha}\mathcal{I}_{l\alpha-n}(1-t)^{n-l\alpha-\frac{1}{2}}=2\pi \ii \mathcal{I}_{l\alpha}\mathcal{I}_{l\alpha-n}(1-t)^{n-l\alpha-\frac{1}{2}}(\tau^{(l)}-\tau^{(l+n N)}).
            \end{equation}
            Together with (\ref{eqn:tau-over-E2}), this gives 
            \begin{equation}
                \frac{\partial H_n^{(l)}}{\partial E_2^{(l)}}=0,
            \end{equation}
            thus by the homogeneity argument, it is a rational function of $t$. 
            \par 
            Proving the explicit formula (\ref{eqn:Hnalpha}) is more complicated. The crucial idea is to allow arbitrary complex values of $l\alpha$ and study the analytical structure of the right-hand sides of (\ref{eqn:H2b}). Since $E_2^{(0)}=0$, by (\ref{eqn:tau-l-nu}), $H_n^{(l)}$ has poles at $l\alpha=k$, $k=0,\ldots, n$. Let us for a moment assume that
            \begin{equation}
            \label{eqn:educated-guess}    H_{n,l\alpha}=\sum_{k=0}^{n-1}\frac{H_{n,k}^{(l)}}{\alpha-k},
            \end{equation}
            where $H_{n,k}^{(l)}$ are polynomials of $t$. Then (\ref{eqn:Hnalpha}) follows by computing the residues at $\alpha=k$ of both hand sides of (\ref{eqn:H2b})\footnote{The assumption that $H_{n,k}^{(l)}$ are polynomials allows to ignore possibility of any nontrivial singularity of the hypergeometric functions, because they can be treated as formal power series.}. The ansatz (\ref{eqn:educated-guess}) can be checked directly for $n=1,2$, and, for larger $n$, proven inductively, using a simple recurrence relation on $H_{n,l\alpha}$. It is convenient to introduce a function
            \begin{equation}
                \tilde{H}_{l,l'}=H_{\frac{l-l'}{N},l\alpha}(1-t)^{-l\alpha}.
            \end{equation}
            Just from its definition and (\ref{eqn:H2b}) follows a Ptolemy-like identity
            \begin{equation}
                \label{eqn:almagest}\tilde{H}_{l_1,l_3}\tilde{H}_{l_2,l_4}=\tilde{H}_{l_1,l_2}\tilde{H}_{l_3,l_4}+\tilde{H}_{l_1,l_4}\tilde{H}_{l_2,l_3},
            \end{equation}
            for any $l_1\equiv l_2\equiv l_3 \equiv l_4 \ ({\rm mod} \ N)$. Then, if we set $l_1=l$, $l_2=l-n$ $l_3=l-n+1$, $l_4=l-1$, (\ref{eqn:almagest}) allows to express $H_{n,l\alpha}$ via $H_{n',l'\alpha}$ with $n'=1,2,n-1,n-2$ and various values of $l'$, thus allowing to prove the induction step. We omit these technical computations.
            \par 
            Now, (\ref{eqn:E2-higher}) and (\ref{eqn:H2b}) imply
            \begin{equation}
                  \label{eqn:II-almost-magic}\mathcal{I}_{l\alpha}\mathcal{I}_{l\alpha-n}(1-t)^{n-l\alpha-\frac{1}{2}}=H_{n,l\alpha}E_2^{(l)}+H_{n,l\alpha}R_{n}^{(l)} f_2^{(l)}(1-t)^{\frac{1}{2}}.
            \end{equation}
            The only missing ingredient is the rational function $R_{n}^{(l)}$. To find it, let us multiply both hand sides by
            \begin{equation}
            \frac{\mathcal{I}_{l\alpha-1}}{\mathcal{I}_{l\alpha}}=-\frac{l\alpha}{2}\frac{f_2^{(l+N)}}{E_{2}^{(l)}}(1-t)^{-\frac{1}{2}}=-\frac{2}{l\alpha}\frac{E_2^{(l)}}{f_2^{(l)}}(1-t)^{-\frac{1}{2}},
            \end{equation}
            we have
            \begin{equation} \label{eqn:almost-magic-n-1-1}
                \mathcal{I}_{l\alpha-1}\mathcal{I}_{l\alpha-n}(1-t)^{n-l\alpha-\frac{1}{2}}=-\frac{l\alpha}{2}H_{n,l\alpha}f_2^{(l+N)}(1-t)^{-\frac{1}{2}}-\frac{2}{l\alpha}H_{n,l\alpha}R_{n}^{(l)} E_2^{(l)}.
            \end{equation}
            At the same time,
            \begin{equation} \label{eqn:almost-magic-n-1-2}
                \mathcal{I}_{l\alpha-1}\mathcal{I}_{l\alpha-n}(1-t)^{n-l\alpha-\frac{1}{2}}=H_{n-1,l\alpha-1}E_2^{(l+N)}+H_{n-1,l\alpha-1}R_{n-1}^{(l+N)} f_2^{(l+N)}(1-t)^{\frac{1}{2}}.
            \end{equation}
            Using (\ref{eqn:E2-higher}) for $n=2$, one can derive
            \begin{equation}
                E_2^{(l+N)}=-E_2^{(l)}-\frac{l^2\alpha^2}{4}\frac{f_2^{(l+N)}}{R_2^{(l)}}(1-t)^{-\frac{1}{2}}.
            \end{equation}
            Substituting it into (\ref{eqn:almost-magic-n-1-2}), and comparing the result with (\ref{eqn:almost-magic-n-1-1}), we ultimately arrive to
            \begin{equation}\label{eqn:Rnl}
                R_n^{(l)}=\frac{\alpha}{2t}\frac{H_{n-1,l\alpha-1}}{H_{n,l\alpha}},
            \end{equation}
            which makes (\ref{eqn:II-almost-magic}) equivalent to (\ref{eqn:magic-II}).

    \section{Explicit formulas and examples}
    
        \subsection{Explicit formulas for the coefficients \texorpdfstring{$\nu_k^{(n)}$}{nu k (n)}}\label{sub-app:nukn}
            In this Appendix, we present explicit formulas for  the coefficients $\nu_k^{(n)}$ appearing in (\ref{eqn:tau-nu}) and (\ref{eqn:tau-l-nu}). These formulas are convenient than one wants to the corrected coupling constant up to few orders of both $q$ and $\varepsilon$ for specific values of $l$ and $N$. The main advantage of this approach in comparison with (\ref{eqn:best-for-tau-mass}) is that the dependence of the masses is concentrated in the coefficients $c_r^{(n)}$ and ${c'}_r^{(n)}$ defined by  
            \begin{equation}
                \sum_{r=0}^{\infty} \frac{c^{(n)}_r}{y^{rN}} =\left(\frac{\tilde{Q}(A y)}{(Ay)^{2N}}\right)^{-\frac{n}{2}}=1+\frac{nT}{2 A^N y^{N}}+\ldots,
            \end{equation}
             
            \begin{equation}
                \sum_{r=0}^{\infty} \frac{c'^{(n)}_r}{y^{rN}} =\left(\left(\frac{\tilde{Q}(Ay)}{(Ay)^{2N}}\right)^{-\frac{1}{2}}-1\right)^{n}=\left(\frac{T}{2 A^N y^N}\right)^n+\ldots      
            \end{equation}
            instead of being somehow encoded in a nontrivial function $V(s)$. 
            All the coefficients above can be expressed via $c_r=c_r^{(1)}$ by
            \begin{equation}
                c^{(n)}_r=\underset{r_1+\ldots+r_n=r}{\sum_{r_1=0}^{\infty}\cdots\sum_{r_n=0}^{\infty}}c_{r_1}\cdots c_{r_n},
            \end{equation}
    
            \begin{equation}
                c'^{(n)}_r=\underset{r_1+\ldots+r_n=r}{\sum_{r_1=1}^{\infty}\cdots \sum_{r_n=1}^{\infty}}c_{r_1}\cdots c_{r_n}.
            \end{equation}
            Clearly,
            \begin{equation}\label{eqn:c-bigo}
                c_r^{(n)}=\mathcal{O}(\varepsilon^{r}),\ {c'}_r^{(n)}=\mathcal{O}(\varepsilon^{r}),\ {c'}_r^{(n)}=\mathcal{O}(\varepsilon^{n}).
            \end{equation}
            Then, after some straightforward algebra, one gets
            \begin{gather}
                % A\sqrt{N}\delta_{l,N}+v_l=-A\sqrt{N}\sum_{n=0}^{\infty}\underset{k_1+\ldots+k_n\equiv l+1 \ \mod N}{\sum_{k_1=1}^{N-1}\cdots \sum_{k_n=1}^{N-1}}\sum_{r=0}^{\infty}(-1)^{k_1+\ldots+k_n}\frac{p_{k_1-1}\cdots p_{k_n-1}}{n!}\nu^{(n)}_{k_1+\ldots+k_n}(\delta p_{N-1}),
                % \\
                \nu^{(n)}_{k}(\delta p_{N-1})=\sum_{n'=0}^{\infty}(-1)^{n'N}\mu^{(n+n')}_{k+n'N}\frac{(\delta p_{N-1})^{n'}}{n'!}\label{eqn:old-nu}
                \\
                \mu^{(n)}_{k}=\sum_{r=0}^{\infty}\sum_{r'=0}^{\infty}\sum_{n''=0}^{\infty}\sum_{n'''=0}^{\infty}\frac{(-1)^{n'''}c^{(n)}_{r}{c'}_{r'}^{(n''+n''')}U^{Nn'''}}{n''!n'''!}\xi_{\frac{k-1}{N}+r+r'+n'''}^{(n+n''+n''')},\\
                \xi_h^{(n)}=(-1)^n\frac{(1-h)_{n-1}}{N}\mathcal{I}_{h-n}(\kappa)U^{-h N}.
            \end{gather}
             In particular, in the first order of $\varepsilon$ we have\footnote{We used the natural guess $\delta p_{N-1}=\mathcal{O}(\varepsilon)$ which was immediately verified.}
            \begin{gather}
                \nu_k^{(n)}=(-1)^n U^{1-k}\frac{\left(1+\frac{1-k}{N}\right)_{n-1}}{N}\mathcal{I}_{\frac{k-1}{N}-n}(\kappa)-\\
                (-1)^nU^{1-k-N}\frac{\left(\frac{1-k}{N}\right)_{n}}{2N}\left(\left(T+2(-1)^N \delta p_{N-1}^{(0)}\right)\mathcal{I}_{\frac{k-1}{N}-n}(\kappa)-T\mathcal{I}_{\frac{k-1}{N}-n+1}(\kappa)\right)+\mathcal{O}(\varepsilon^2).
            \end{gather}
            From $\nu_0^{(0)}=-1$ we have
            \begin{equation}
                T+2(-1)^N \delta p_{N-1}^{(0)}=T\frac{\mathcal{I}_{1-\frac{1}{N}}(\kappa)}{-\mathcal{I}_{\frac{1}{N}}(\kappa)},
            \end{equation}
            so
            \begin{gather}
                \nu_k^{(n)}=(-1)^n U^{1-k}\frac{\left(1+\frac{1-k}{N}\right)_{n-1}}{N}\mathcal{I}_{\frac{k-1}{N}-n}(\kappa)+\\
                \frac{T}{A^N}(-1)^nU^{1-k-N}\frac{\left(\frac{1-k}{N}\right)_{n}}{2N}\left(\mathcal{I}_{\frac{k-1}{N}-n+1}(\kappa)-\mathcal{I}_{\frac{k-1}{N}-n}(\kappa)\frac{\mathcal{I}_{1-\frac{1}{N}}(\kappa)}{\mathcal{I}_{-\frac{1}{N}}(\kappa)}\right)+\mathcal{O}(\varepsilon^2).
            \end{gather}
            \par
            This gives in the first order (using the formulas from Appendix \ref{app:modularprop})
            \begin{eqnarray}
                \tau^{(l)}=(\tau^{(l)})_{\varepsilon=0}+\frac{(-1)^N}{2\pi\ii}  \frac{T}{A^N} (1-t)^{\frac{1}{2}-\frac{l}{N}}\frac{\mathcal{I}_{-\frac{1}{N}}^{N}(\kappa)}{\mathcal{I}_{-\frac{l}{N}}^{2}(\kappa)}+\mathcal{O}(\varepsilon^2
                ) \\ =(\tau^{(l)})_{\varepsilon=0}+\frac{(-1)^N}{2\pi\ii} \frac{T}{A^N} \frac{f_2^{\frac{N}{2}}(q_{\rm IR})}{f_2^{(l)}(q_{\rm IR}^{(l)})}+\mathcal{O}(\varepsilon^2
                )
            \end{eqnarray}
            % where
            % \begin{equation}
            %     t=\frac{4}{\kappa^2}=\frac{4}{q+\frac{1}{q}+2}
            % \end{equation}

            We note that the computational power needed to compute the coefficients $\nu$ using (\ref{eqn:old-nu}) increases drastically than higher orders of $\varepsilon$ are taken into account. Moreover, integration of the resulting nonlinear combinations of hypergeometric functions without the ideas of Subsections \ref{ssec:mod-obs}  \ref{ssec:mod-obs} can be done only with a fair bit of luck.

        \subsection{Coefficients \texorpdfstring{$\Phi_{l}^{(n)}$}{Phi-l-n} and \texorpdfstring{$\Psi_{l}^{(n)}$}{Psi-l-(n)}}   \label{app:phipsi}
            Combining (\ref{eqn:nu-def}) and (\ref{eqn:nuhat-def}), we get
            \begin{eqnarray}
                 \hat{\nu}_{l+1+(n-1)N}^{(n)}(\delta p_{N-1})=- \sum_{l'\equiv l \, \mod N}{\rm Res}_{s=0}\left(s^{ N-l+l'-1}U^{-l'} \frac{V_{l'}^{(n)}(\delta p_{N-1})}{l'}\right)\mathcal{I}_{-\frac{l}{N}}\mathcal{I}_{\frac{l'}{N}}= \nonumber
                 \\
                 (-1)^{nN}
                 U^{-l}\frac{\partial^{n}}{\partial \delta p_{N-1}^n}\sum_{f\in\mathbb{Z}}{\rm Res}_{S=0}\left(S^{-f-1}U^{-(f-1)N} \frac{V(\delta p_{N-1},S^{\alpha})^{l\alpha-f+1}}{N(l\alpha-f+1)}\right)\mathcal{I}_{l\alpha}\mathcal{I}_{-l\alpha+f-1}.\label{eqn:nuhat}
            \end{eqnarray}
            Here we have put $l'=l+(f-1) N$, introduced a new variable $S=s^{-N}$, substituted (\ref{eqn:Vln}) and used the fact that in any finite order of $\varepsilon$, 
            the expression under the residue is a Laurent polynomial of $s$, so the residue at $S=\infty$ is equal to the one at $S=0$ up to a sign. By (\ref{eqn:Vdef}), 
            \begin{equation}\label{eqn:VviaS}
                V(\delta p_{N-1},S^{\alpha})=1-(-1)^{N}\delta p_{N-1} U^{-N}G_1(S^{\alpha})-U^{-N}\left(S^{-1}-U^N\right)(G_1(S^{\alpha})-1),
            \end{equation}
            where, by (\ref{eqn:G1-Def}),
            \begin{equation}
                G_1(S^\alpha)=\left(1-\frac{T}{A^N} S+\frac{T'}{A^{2N}}S^2\right)^{-\frac{1}{2}}.
            \end{equation}
            \par We note that at $S\to 0$, $G_1(S^{\alpha})-1=\frac{T}{2A^N}S+\mathcal{O}(S^2)$, so the right-hand side of (\ref{eqn:VviaS}) is actually regular at $S=0$. This implies that in (\ref{eqn:nuhat}) we can leave only the nonnegative values of $f$. Using (\ref{eqn:magic-II-minus}), we get the desired form 
            \begin{equation}
                   \hat{\nu}_{l+1+(n-1)N}^{(n)}=U^{-l}\left({\Phi}^{(n)}_{l}E_2^{(l)}+{\Psi}^{(n)}_{l}f_2^{(l)}(1-t)^{\frac{1}{2}}\right)
            \end{equation}
            with 
            \begin{equation}\label{eqn:Phi}
                \Phi_l^{(n)}=(-1)^{nN}\frac{\partial^{n}}{\partial \delta p_{N-1}^n}\sum_{f=0}^{\infty}{\rm Res}_{S=0}\left(S^{-f-1}U^{-(f-1)N} \frac{V(\delta p_{N-1},S^{\alpha})^{l\alpha-f+1}}{N(l\alpha-f+1)}\right)H_{f,l\alpha},
            \end{equation}
            \begin{equation}\label{eqn:Psi}
                \Psi_l^{(n)}=\frac{l\alpha}{2}(-1)^{nN}\frac{\partial^{n}}{\partial \delta p_{N-1}^n}\sum_{f=0}^{\infty}{\rm Res}_{S=0}\left(S^{-f-1}U^{-(f-1)N} \frac{V(\delta p_{N-1},S^{\alpha})^{l\alpha-f+1}}{N(l\alpha-f+1)}\right)H_{f-1,l\alpha-1}.
            \end{equation}
            Analogously, we write
            \begin{eqnarray}
                  \hat{\nu}_{nN-l+1}^{(n)}(\delta p_{N-1})=- \sum_{l'\equiv -l \, \mod N}{\rm Res}_{s=0}\left(s^{ l+l'-1}U^{-l'} \frac{V_{l'}^{(n)}(\delta p_{N-1},s)}{l'}\right)\mathcal{I}_{\frac{l}{N}-1}\mathcal{I}_{\frac{l'}{N}}=
                  \\
                  (-1)^{nN}U^{l}\frac{\partial^{n}}{\partial \delta p_{N-1}^n}{\rm Res}_{S=0} S^{-f-1}U^{-fN}\frac{V(\delta p_{N-1},S^{\alpha})^{-l\alpha-f}}{N(-l\alpha-f)}(1-t)^{-l\alpha-\frac{1}{2}}\mathcal{I}_{l\alpha}\mathcal{I}_{l\alpha+f},
            \end{eqnarray}
            where this time we set $l'=fN-l$ and have used (\ref{eqn:prop1}). Using (\ref{eqn:magic-II}), we get
            \begin{equation}
                   \hat{\nu}_{nN-l+1}^{(n)}=U^{l}\left(\widetilde{\Phi}^{(n)}_{l}E_2^{(l)}+\widetilde{\Psi}^{(n)}_{l}f_2^{(l)}(1-t)^{\frac{1}{2}}\right)
            \end{equation}
            with 
            \begin{equation} \label{eqn:tildephi}
                \widetilde{\Phi}_l^{(n)}=(-1)^{nN}\frac{\partial^{n}}{\partial \delta p_{N-1}^n}\sum_{f=0}^{\infty}{\rm Res}_{S=0}\left(S^{-f}U^{-fN} \frac{V(\delta p_{N-1},S^{\alpha})^{-l\alpha-f}}{N(-l\alpha-f)}\right)(1-t)^{f}H_{-f,l\alpha},
            \end{equation}
            \begin{equation} \label{eqn:tildepsi}
                \widetilde{\Psi}_l^{(n)}=(-1)^{nN}\frac{l\alpha}{2}\frac{\partial^{n}}{\partial \delta p_{N-1}^n}\sum_{f=0}^{\infty}{\rm Res}_{S=0}\left(S^{-f}U^{-fN} \frac{V(\delta p_{N-1},S^{\alpha})^{-l\alpha-f}}{N(-l\alpha-f)}\right)(1-t)^{f}H_{-f-1,l\alpha-1}.
            \end{equation}
            
        \subsection{Examples of computations} \label{app:examplestau}
            Let us present some examples of order-by-order evaluation of $\tau^{(l)}_{\bm m}$ from (\ref{eqn:tau-nu}) with (\ref{eqn:bound1}). 
            \par For $N=4$ we have two coupling constants,
            \begin{gather}
                \tau_{\bm m}^{(1)}=\tau^{(1)}-\frac{T}{2\pi\ii A^N}\left(1-\frac{3}{8}q-\frac{69}{512}q^2-\frac{303}{4096}q^3+\mathcal{O}(q^4)\right)+\mathcal{O}(\varepsilon^2)=
                \\
                \tau^{(1)}-\frac{T}{2\pi\ii A^N}\left(1+24q_{\rm IR}+24q_{\rm IR}^2+96q_{\rm IR}^3+\mathcal{O}(q^4)\right)+\mathcal{O}(\varepsilon^2)
                \nonumber
            \end{gather}
            and        
            \begin{gather}
                \tau_{\bm m}^{(2)}=\tau^{(2)}-\frac{T}{2\pi\ii A^N}\left(1-\frac{1}{4}q-\frac{25}{256}q^2-\frac{29}{512}q^3+\mathcal{O}(q^4)\right)+\mathcal{O}(\varepsilon^2),
            \end{gather}
            while for $N=6$ there are three constants,
            \begin{gather}
                \tau_{\bm m}^{(1)}=\tau^{(1)}-\frac{T}{2\pi\ii A^N}\left(1-\frac{5}{9}q-\frac{185}{1296}q^2-\frac{440}{5461}q^3+\mathcal{O}(q^4)\right)+\mathcal{O}(\varepsilon^2)=
                \\
                \tau^{(1)}-\frac{T}{2\pi\ii A^N}\left(1+240q_{\rm IR}+55440q_{\rm IR}^2+12793920 q_{\rm IR}^3+\mathcal{O}(q^4)\right)+\mathcal{O}(\varepsilon^2),
                \nonumber
            \end{gather}
    
            \begin{gather}
                \tau_{\bm m}^{(2)}=\tau^{(2)}-\frac{T}{2\pi\ii A^N}\left(1-\frac{7}{18}q-\frac{319}{2592}q^2-\frac{27413}{419904}q^3+\mathcal{O}(q^4)\right)+\mathcal{O}(\varepsilon^2),
            \end{gather}
            and
            
            \begin{gather}
                \tau_{\bm m}^{(3)}=\tau^{(3)}-\frac{T}{2\pi\ii A^N}\left(1-\frac{1}{3}q-\frac{47}{432}q^2-\frac{257}{4374}q^3+\mathcal{O}(q^4)\right)+\mathcal{O}(\varepsilon^2).
            \end{gather}
            Here $q_{\rm IR}=\ee^{2\pi\ii \tau^{(1)}}$. These results agree with \cite{Lerda} up to different parametrisation of the mass parameters.


\begin{thebibliography}{10}

        \bibitem{SW1} N. Seiberg and E. Witten, \textit{Monopole Condensation, And Confinement In N=2 Supersymmetric Yang-Mills Theory}, Nucl. Phys. B 426 (1994) 19, \href{https://arxiv.org/abs/hep-th/9407087}{arXiv:hep-th/9407087}.
        
        \bibitem{SW2} N. Seiberg and E. Witten, \textit{Monopoles, Duality and Chiral Symmetry Breaking in N=2 Supersymmetric QCD}, Nucl. Phys. B 431 (1994) 484, \href{https://arxiv.org/abs/hep-th/9408099}{arXiv:hep-th/9408099}.

        \bibitem{AGT} L.F. Alday, D. Gaiotto, Y. Tachikawa, \textit{Liouville Correlation Functions from Fourdimensional Gauge Theories}, Lett. Math. Phys. 91 (2010) 167-197, \href{https://arxiv.org/abs/0906.3219}{arXiv:0906.3219  [hep-th]}.

        \bibitem{Pesandoetal} M. Billo, M. Frau, L. Gallot, A. Lerda, and I. Pesando, \textit{Deformed N=2 theories, generalized recursion relations and S-duality}, J. High Energ. Phys. 1304 (2013) 039, \href{https://arxiv.org/abs/1302.0686}{arXiv:1302.0686 [hep-th]}.

        \bibitem{Pesandoetal2} M. Billo, M. Frau, L. Gallot, A. Lerda, and I. Pesando, \textit{Modular anomaly equation, heat kernel and S-duality in N = 2 theories}, J. High Energ. Phys. 1311 (2013) 123, \href{https://arxiv.org/abs/1307.6648}{arXiv:1307.6648 [hep-th]}.

        \bibitem{Moralesetal} M. Billò, M. Frau, F. Fucito, A. Lerda, J. F. Morales, \textit{Resumming instantons in N=2* theories with arbitrary gauge groups}, in proceedings of the 14th Marcel Grossmann Meeting (MG14) Rome, Italy, July 12-18, 2015, \href{https://arxiv.org/abs/1602.00273}{arXiv:1602.00273}.
        
        \bibitem{ADE} M. Billo, M. Frau, F. Fucito, A. Lerda and J. F. Morales, \textit{S-duality and the prepotential in N=2* theories (I): the ADE algebras}, J. High Energ. Phys. 11 (2015) 024, \href{https://arxiv.org/abs/1507.07709}{arXiv:1507.07709 [hep-th]}.

        \bibitem{nonsimplaced}  M. Billo, M. Frau, F. Fucito, A. Lerda and J. F. Morales, \textit{S-duality and the prepotential in N=2* theories (II): the nonsimply laced algebras}, J. High Energ. Phys. 11 (2015) 026, \href{https://arxiv.org/abs/1507.08027}{arXiv:1507.08027 [hep-th]}.
        
        \bibitem{MinahanNemeschansky}J. A. Minahan and D. Nemeschansky,\textit{ N=2 superYang-Mills and subgroups of SL(2,Z)}, Nucl. Phys. B468 (1996) 7284 \href{https://arxiv.org/abs/hep-th/9601059v2}{arXiv:hep-th/9601059}.

        \bibitem{Frauetal} S. K. Ashok, M. Billo, E. Dell’Aquila, M. Frau, A. Lerda and M. Raman,\textit{ Modular anomaly equations and S-duality in N = 2 conformal SQCD}, J. High Energ. Phys. 10 (2015) 091, \href{https://arxiv.org/abs/1507.07476}{arXiv:1507.07476}.
        
        \bibitem{Billoetal} M. Billo, M. Frau, F. Fucito, L. Giacone, A. Lerda, J. F. Morales, and D. R. Pacifici, \textit{Non-perturbative gauge/gravity correspondence in N=2 theories}, J. High Energ. Phys. 08 (2012) 166, \href{https://arxiv.org/abs/1206.3914v1}{arXiv:1206.3914v1 [hep-th]}.

        \bibitem{Lerda} S. K. Ashok, E. Dell’Aquila, A. Lerda and M. Raman, \textit{S-duality, triangle groups and modular anomalies in N=2 SQCD}, J. High Energ. Phys. 04 (2016) 118, \href{https://arxiv.org/abs/1601.01827}{arXiv: 1601.01827}.

         \bibitem{NekrasovOkunkov} Nikita Nekrasov, Andrei Okounkov, \textit{Seiberg-Witten theory and random partitions}, Prog.Math. 244 (2003) 525-596 \href{https://arxiv.org/abs/hep-th/0306238v2}{arXiv:hep-th/0306238v2}

        \bibitem{ArgyresPelland} P. C. Argyres and S. Pelland, \textit{Comparing instanton contributions with exact results in N=2 supersymmetric scale invariant theories}, J. High Energ. Phys. 0003 (2000) 014, \href{https://arxiv.org/abs/hep-th/9911255}{hep-th/9911255}   

        \bibitem{BykovSysoeva} Aleksei Bykov, Ekaterina Sysoeva, \textit{Zamolodchikov recurrence relation and modular properties of effective coupling in $\mathcal{N}=2$ SQCD}, 	\href{https://arxiv.org/abs/2507.20876}{arXiv:2507.20876 [hep-th]}.

        \bibitem{Nekrasov} Nikita Nekrasov, \textit{Seiberg-Witten Prepotential from Instanton Counting}, Advances in Theoretical and Mathematical Physics. 7(5) (2003) 831-864, \href{https://arxiv.org/abs/hep-th/0206161}{arXiv:hep-th/0206161}.
    
        \bibitem{SeibergWitten}  N. Seiberg, E. Witten, \textit{Electric - magnetic duality, monopole condensation, and confinement in N=2 supersymmetric Yang-Mills theory}, Nucl. Phys. B. 426(1) (1994) 19–52, \href{https://arxiv.org/abs/hep-th/9407087v1}{arXiv:hep-th/9407087v1}

        \bibitem{ArgyresBuchel} P. C. Argyres and A. Buchel, \textit{The Nonperturbative gauge coupling of N=2 supersymmetric theories}, Phys.Lett. B442 (1998) 180–184, \href{https://arxiv.org/abs/hep-th/9806234v2}{arXiv:hep-th/9806234v2}.

        \bibitem{NIST} National Institute of Standards and Technology \href{https://dlmf.nist.gov/15.5}{website}.

        \bibitem{Doran} Charles F. Doran, Terry Gannon, Hossein Movasati, Khosro Monsef Shokri, \textit{Automorphic forms for triangle groups}, Commun. Num. Theor. Phys. 07 (2013) 689--737, \href{https://arxiv.org/abs/1307.4372}{	arXiv:1307.4372 [math.NT]}.

        \bibitem{Zagier} ] J. H. Bruinier, G. van der Geer, G. Harder, D. Zagier, \textit{The 1-2-3 of Modular Forms.}, Springer, Berlin (2008) ISBN: 978-3-540-74117-6.

    \end{thebibliography}
\end{document}